\pdfoutput=1
\documentclass[%
prx,twocolumn,
notitlepage,
superscriptaddress,
 amsmath,amssymb,
 aps,
prx,
floatfix,
longbibliography
]{revtex4-1}
\usepackage[T1]{fontenc}
\usepackage{float}
\usepackage{color}
\usepackage{amsmath}
\usepackage{amssymb}
\usepackage{bm}
\usepackage{graphicx}
\usepackage[usenames,dvipsnames]{xcolor}
\usepackage[%
  colorlinks=true,
  urlcolor=blue,
  linkcolor=blue,
  citecolor=blue
]{hyperref}

\usepackage{array}
\definecolor{bleu}{rgb}{0.16,0.2.5,0.36}

\newcommand{\EDF}{\textbf{Fig.}}%to use for submission to journal
\usepackage{lipsum} % for dummy text
\begin{document}
\pdfoutput=1
%\title{Non-orientable  order: from twisted sheets to frustrated matter}

%\title{Ideal shock absorbers harnessing yield buckling}
\title{Leveraging yield buckling to achieve ideal shock absorbers}
%\title{Non-orientable Order}
\author{Wenfeng Liu}
\affiliation{Institute of Physics, Universiteit van Amsterdam, 1098 XH Amsterdam, The Netherlands}
\author{Shahram Janbaz}
\affiliation{Institute of Physics, Universiteit van Amsterdam, 1098 XH Amsterdam, The Netherlands}
\author{David Dykstra}
\affiliation{Institute of Physics, Universiteit van Amsterdam, 1098 XH Amsterdam, The Netherlands}
\author{Bernard Ennis}
\affiliation{Tata Steel, 1970 CA IJmuiden, The Netherlands}
\author{Corentin Coulais}
\affiliation{Institute of Physics, Universiteit van Amsterdam, 1098 XH Amsterdam, The Netherlands}

\begin{abstract}
%motivation
The ideal shock absorber combines high stiffness with high energy absorption whilst retaining structural integrity after impact and is scalable for industrial production. 
%state-of-the-art
So far no structure meets all of these criteria.
%Although many promising structures have been previously presented in the literature combining one or more of these features, none have yet to meet them all.
%breakthrough
Here, we introduce a special occurrence of plastic buckling as a design concept for mechanical metamaterials that combine all the elements required of an ideal shock absorber. 
%In this paper we design, create and validate mechanical metamaterials which combine all the elements required of an ideal shock absorber. 
%Methods
By striking a balance between plastic deformation and buckling, which we term yield buckling, these metamaterials exhibit sequential, maximally dissipative collapse combined with high strength and the preservation of structural integrity.
%Impact
Unlike existing structures, this design paradigm is applicable to all elastoplastic materials at any lengthscale and hence will lead to a new generation of shock absorbers with enhanced safety and sustainabilty 
%opens avenues for a new generation of shock absorbers 
in a myriad of high-tech applications.
\end{abstract}

\date{\today}
\maketitle
Shock absorption is a crucial function designed to protect objects, structures, or living organisms from the damaging effects of sudden impacts or vibrations.
It acts as a buffer by absorbing and dissipating the energy generated during such events.
In mechanical systems, shock absorption helps prevent excessive stress, wear, and damage by reducing the intensity and duration of the impact.
%car example: needed? 
A car crash is an example of a shock with potentially devastating consequences. 
The impact of the crash needs to be mitigated to protect the contents and, most importantly, the passengers in the vehicle. 
%define ideal shock absorber
The ideal shock-absorbing material is defined by the following characteristics. First, it should exhibit high stiffness and strength in order to act as a load-bearing structure. Second, 
%that exhibits exceptional energy absorption capabilities during such high-impact events. 
it should exhibit progressive collapse with a constant deceleration that maximally absorbs impact energy and able to sustain multiple impacts.
Third, for any practical application, one should be able to mass-produce it at the industrial scale.%high-tech applications such as vehicles, aircraft, and seismic protection.
%At the same time, it should exhibit high stiffness and strength in order to act as a load-bearing structure in 

%, controlled deformation, progressive collapse, and effective energy dissipation to mitigate the forces and protect occupants or structures from severe damage during crashes or collisions (Fig.~\ref{fig:1} A red).

%Paragraph before: you need negative post-buckling stiffness to achieve sequential buckling. But so far, this has primarily been achieved in elastomeric structures (snap-through buckling of arches\cite{Katia,EML} and nonlinear beams\cite{DiscontinuousPRL,AdvMatFunc}).

\begin{figure}[b!]
    \centering
\includegraphics[width=1.0\linewidth]{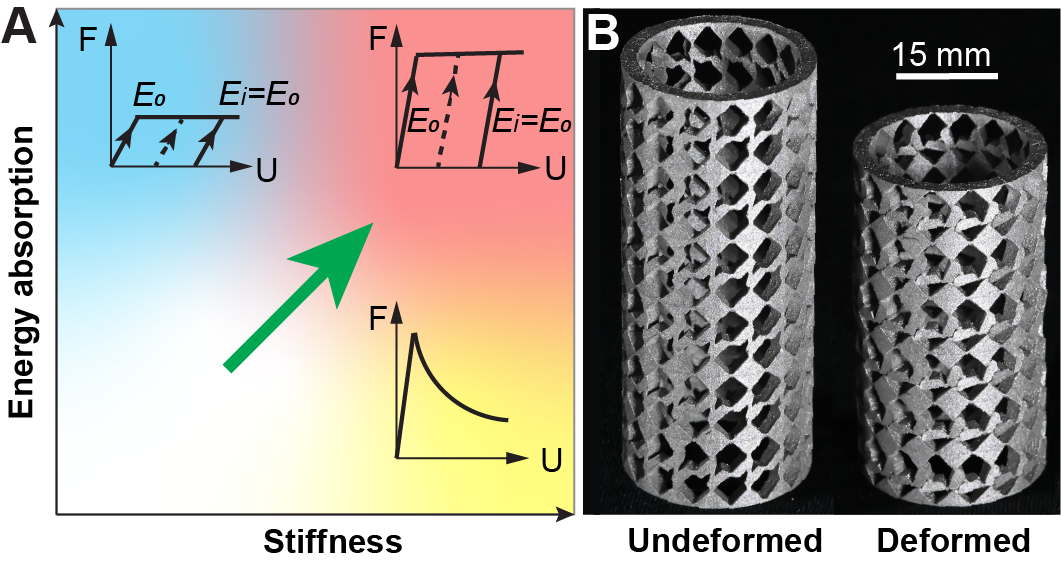}
\caption{\textbf{Ideal shock absorbers.} \textbf{A,} (bottom-right) stretching-dominated materials are stiff but do not exhibit a force plateau. Bending-dominated materials exhibit a force plateau a low stiffness and are reusable (top-left). Ideal shock absorbers exhibit a high stiffness before impact for load bearing, show a stable plateau during impact for maximum energy absorption, are reusable multiple times while retaining their initial stiffness (top-right) and can be mass-manufactured. \textbf{B,} a cylindrical metamaterial with multiple layers strip mode designed for such ideal shock absorption enabled by yield buckling.}
    \label{fig:1}
\end{figure}

So far, no metamaterial or any other structure simultaneously meets all of the criteria of ideal shock absorbers. 
First, scalable structures such as traditional crash cans~\cite{ABRAMOWICZ1984263,WIERZBICKI1983157, dipaolo2006study, wierzbicki1983crushing} and stretching-dominated metamaterials~\cite{deshpande2001foam,zheng2014ultralight} 
%are precisely engineered materials that enable 
exhibit high specific stiffness and strength. Yet once they start to buckle, they collapse catastrophically
%, which leads to large oscillations in collapse stress after buckling
~\cite{deshpande2001foam, deshpande2003energy}. This collapse limits the efficiency of the shock absorption (Fig.~\ref{fig:1}A-bottom right). 
Second, traditional foams~\cite{gibson2003cellular, ashby2000metal}, bending-dominated metamaterials~\cite{babaee20133d, bauer2021tensegrity} and metamaterials that exploit snap-through buckling~\cite{florijn2014programmable, shan2015multistable, RESTREPO201552, rafsanjani2015snapping} or Euler buckling with negative stiffness~\cite{chen2021reusable, coulais2015discontinuous, lubbers2017nonlinear} exhibit progressive collapse by smooth~\cite{bauer2021tensegrity,babaee20133d} or sequential deformations~\cite{florijn2014programmable, shan2015multistable, RESTREPO201552, frenzel2016tailored,coulais2018multi,rafsanjani2015snapping}. This progressive collapse is more efficient as it distributes the absorption of energy evenly throughout the impact stroke and can be reused multiple times. 
%Buckling with negative stiffness enables a progressive and reusable energy absorption, as previously shown in bistable structures, and nonlinear elastomeric beams 
Yet these metamaterials rely on low-stiffness components---bending and snapping arches or low-stiffness materials---which limit their load-bearing capacity (Fig.~\ref{fig:1}A-top left).
%
%Third, metal microlattices~\cite{tancogne2016additively}---that exploit thick components which plastify without buckling---and nanolattices~\cite{meza2014strong, evans2010concepts}---that exploit defect-free nanoscale components and shell buckling---exhibit high stiffness and a force plateau simultaneously. Yet these metamaterials are complex three-dimensional architectures that can only be made by additive manufacturing and hence are hard to produce at the industrial scale.
Third, nanolattices~\cite{meza2014strong, evans2010concepts} which exploit defect-free nanoscale building blocks and shell buckling can combine high stiffness, a force plateau and are reusable. 
Yet these metamaterials %are only reusable at low density
rely %on components which are restricted to the nanometer scale and 
on complex three-dimensional nanoscale components, which can be produced solely by additive manufacturing. Therefore, it would be extremely challenging to deploy these nanolattices in practical applications at full scale.%, both in size and quantity.

\begin{figure*}[t!]
\centering
\includegraphics[width=1\linewidth]{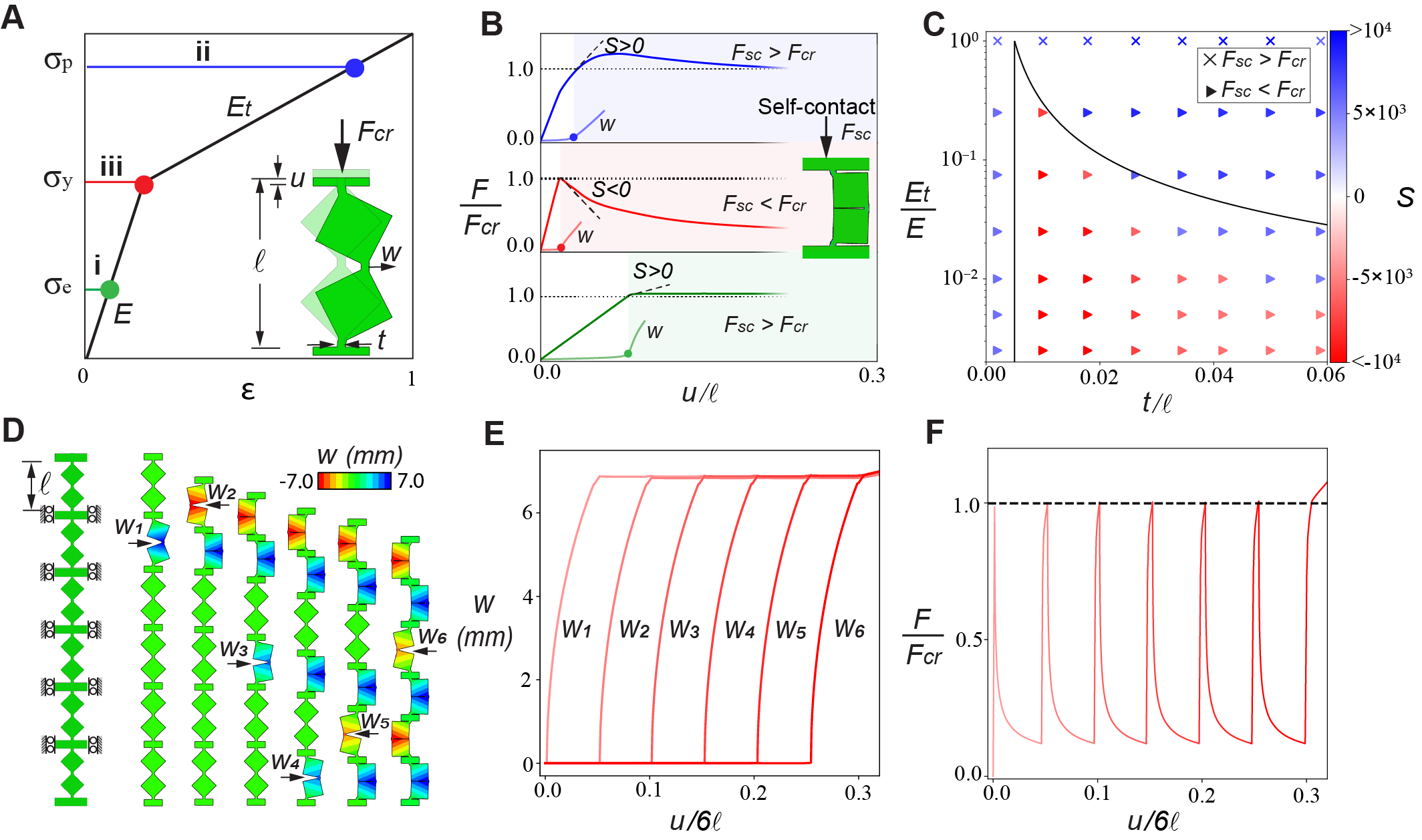}
\caption{\textbf{Yield buckling.} \textbf{A.}  Stress $\sigma$ vs. strain $\varepsilon$ for a bilinear elastoplastic model with Young modulus $E$, yield stress $\sigma_y$ and tangent modulus $E_t$. 
Inset: unit cell consisting of a pair of rotating squares and connecting ligaments buckling at a load $F_{cr}$.
The green (before yielding), red (at yielding), and blue (after yielding) markers denote three stress states of the ligaments when the unit starts to buckle. 
%Three regions of buckling exist for the elastoplastic ligaments with elastic modulus, $E$, yield stress, $\sigma_{y}$, and tangent modulus, $E_{t}$: buckle before plasticity (elastic buckling, green dot), buckle in plasticity (plastic buckling, blue dot), and buckle at start of plasticity (yield buckling, red dot).
\textbf{B.} %For the cases of elastic (green) and plastic buckling(blue), the unit shows positive stiffness after buckling. Only yield buckling shows negative stiffness and monotonic force drop until self-contact (red). 
Force (thick lines) and lateral deflection (thin lines) vs. displacement in the elastic (green), plastic (blue), and yield (red) buckling regimes from finite elements simulations (see SI for details). $F_{cr}$ is the critical load at buckling and $S$ is the slope right after the onset of buckling.
Inset: unit cell at self-contact reached at a load $F_{sc}$.
\textbf{C.} 
Post-buckling stiffness $S$ vs. aspect ratio of the unit cell $t/\ell$ and ratio between tangent and Young's moduli $E_{t}/E$. 
%Yield buckling is
%The loads in the yield buckling regime are always below the initial buckling load 
The red triangles denote the yield buckling regime defined by $S<0$ and $F_{sc}<F_{cr}$.
\textbf{D, E, F} A six-step sequential yield buckling is achieved in a six-layer structure with sliding constraints on its side. \textbf{D.} Snapshots at each buckling step. The colors denote the horizontal displacement field.
%The multi-layer structure shows the same stiffness and strength in each step of buckling. The post-buckling load is always lower than the initial buckling load enabling the next step of buckling can be only triggered by the self-contact.
\textbf{E.} Lateral deflection of the center of each ligament $w_{i}$ vs. compressive stroke $u/6\ell$. \textbf{F.} Load vs. compressive stroke $u/6\ell$. The reaction load is normalized by the initial yield buckling load, $F_{cr}=\sigma_{y}t$.
}
\label{fig:2}
\end{figure*}
Here we introduce a distinctive mechanism to engineer ideal shock absorbers (Fig.~\ref{fig:1}A-top right), which we term ``yield buckling''. %We create metamaterials made from elastoplastic unit cells, that are both stiff prior to buckling and exhibit yield buckling. 
We first show that the regime of yield buckling can arise in any elastoplastic structure made of ligaments that act as plastic hinges and of rigid elements that rotate. In such a regime, the load decreases right at the onset of buckling and remains smaller than the buckling load until the rotating elements reach self-contact. We then harness yield buckling to design metamaterials that have the following attributes: they are stiff and strong, exhibit a progressive collapse that is maximally dissipative and can withstand several impacts while retaining their initial stiffness (Fig.~\ref{fig:1}B). Last but not least, our metamaterials can be mass-manufactured at any scale.
%has done been done with high-stiffness materials such as metals. 
%Here, we achieve buckling with negative stiffness can naturally emerge in elastoplastic structures with high stiffness.

Consider the buckling behavior of a pair of squares of size $\ell$ connected by an elastoplastic ligament of thickness $t$ with Young's modulus $E$, tangent modulus $E_t$ and yield stress $\sigma_y$ (Fig.~\ref{fig:2}A). When compressed along its major axis, the unit exhibits high stiffness before buckling, the ligament undergoes both elastic and plastic deformations, and the pair of squares will start to rotate due to a buckling instability. There are three regimes (Fig.~\ref{fig:2}AB and see also Appendix A).
(i) elastic buckling (green): when the aspect ratio $t/\ell$ of the unit cell is sufficiently small or the yield stress $\sigma_{y}/E$ is sufficiently large, it will start to deflect prior to plastic deformation. As a result, the post-buckling stiffness will remain positive leading to a continuing rise in force \cite{cedolin2010stability, coulais2015discontinuous}; (ii) plastic buckling (blue): when the aspect ratio $t/\ell$ is sufficiently large or the moduli ratio, $E_{t}/E$, is sufficiently large, it will no longer deflect elastically and buckling occurs after the onset of plastic deformation, resulting in the load continuing to rise after buckling \cite{shanley1947inelastic, cedolin2010stability}; (iii) yield buckling (red): when the aspect ratio of the column and plasticity $\sigma_{y}/E$, $E_{t}/E$ of the material are delicately balanced, then buckling occurs \emph{precisely at} the yield point. This results in a sharp load decrease concomitant with buckling. Regime (iii) is of particular interest because the drop in load at buckling guarantees that the load will always remain below the initial buckling load until the unit cell reaches self-contact---this will be crucial to achieving sequential buckling steps later on.
%Three scenarios are likely: 
%the hinge remains in the elastic regime
%at the onset of this instability, the force may increase with a positive stiffness elastic and the buckling (green) and plastic buckling (blue), or decrease until the self-contact of the two squares occurs, leading to yield buckling (Fig.~\ref{fig:2}B red curve). The emergence of yield buckling is attributed to the highly asymmetric stress distribution across the ligament. The compressed side of the ligament undergoes plastic loading with a tangent modulus $E_t$. While the tensile side experiences elastic unloading with an elastic modulus, $E \gg E_t$ (Fig.~\ref{fig:7}F). This unloading predominates, resulting in a reduction of the total reaction force. Based on the stiffness at buckling and the post-buckling load, we define yield buckling as the buckling with a negative buckling stiffness, and the post-buckling load is lower than the initial buckling load before self-contact(Fig.~\ref{fig:2}B red).
%In order to identify the various regime and pinpoint their origin, 

\begin{figure*}[t!]
\centering
\includegraphics[width=1.0\linewidth]{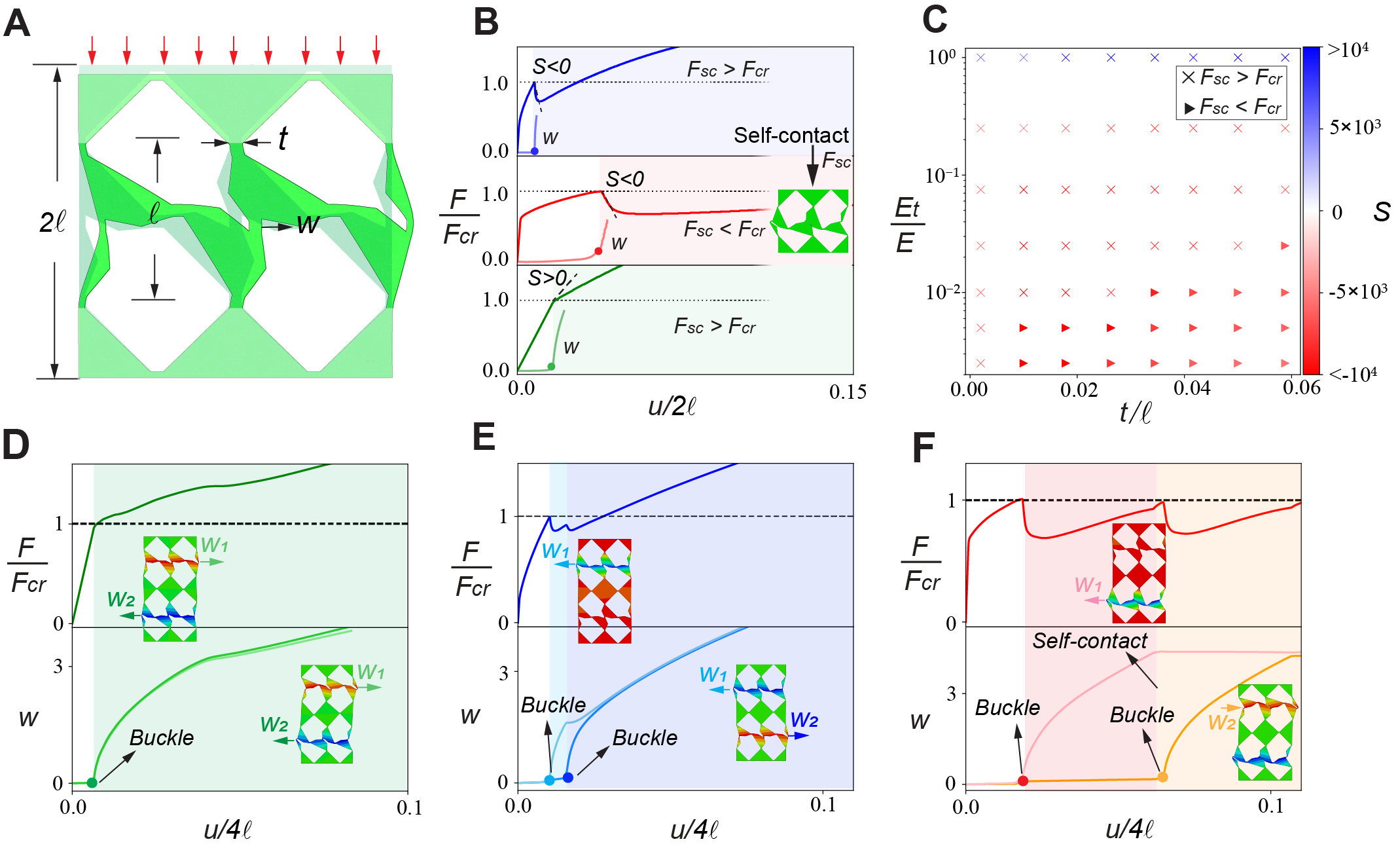}
\caption{\textbf{Yield buckling in a metamaterial with line modes.} \textbf{A.} Snapshot of a finite element simulation of a metamaterial unit cell buckles along lines under compression. 
\textbf{B.} 
Force (thick lines) and lateral deflection (thin lines) vs. displacement in the elastic (green), plastic (blue), and yield (red) buckling regimes from finite elements simulations (see Appendix B for details). $F_{cr}$ is the critical load at buckling and $S$ is the slope right after the onset of buckling.
Inset: unit cell at self-contact reached at a load $F_{sc}$.
%The extra horizontal ligaments expand the region of negative buckling stiffness in the line-mode unit and avoid large force drop in the post-buckling. Three types of line-mode buckling are distinguished here based on the post-buckling stiffness and the post-buckling load before self-contact: buckle with a positive stiffness and a post-buckling load higher than the initial buckling load (elastic buckling); buckle with a negative stiffness and a post-buckling load higher than the initial buckling load (plastic buckling); buckle with a negative stiffness and the post-buckling load always lower than the initial buckling load (yield buckling). 
\textbf{C.} 
Post-buckling stiffness $S$ vs. aspect ratio of the unit cell $t/\ell$ and ratio between tangent and Young's moduli $E_{t}/E$. 
%Yield buckling is
%The loads in the yield buckling regime are always below the initial buckling load 
The red triangles denote the yield buckling regime defined by $S<0$ and $F_{sc}<F_{cr}$. 
%We numerically prove these in a line-mode unit with different ratios of $t/l$, $E_{t}/E$, by measuring the post-buckling stiffness and the self-contact load of the unit. The result shows only elastic buckling in the unit shows positive stiffness at the buckling. The post-buckling stiffness far from the buckling point depends on the tangent modulus and a low tangent modulus can make the post-buckling load always lower the buckling load. 
%
\textbf{DEF.} Force (top) and deflection of the central ligament (bottom) in a finite element simulation of two unit cells in series in the elastic (\textbf{D}), plastic (\textbf{E}) and yield buckling (\textbf{F}) regimes.
%a two-layer line-mode structure is used to describe the simultaneous buckling in elastic buckling,  incomplete sequential buckling in plastic buckling, and complete sequential buckling in yield buckling. 
%
%\textbf{GHI.} Buckling mode vs. aspect ratio $t/\ell$ and metamaterials size $N$ in the elastic (\textbf{G}), plastic (\textbf{H}) and yield buckling (\textbf{I}) regimes from a finite element analysis (see SI for details). Grey squares denote a global counter rotating-square mode~\cite{resch1965geometrical,coulais2018characteristic,czajkowski2022conformal} in a single step (see insets panels \textbf{GH}), green and blue circles denote simultaneous buckling of all line modes (see insets in panels \textbf{GH}), red circles denote sequential buckling of the lines modes (see insets panels \textbf{I}) and grey diamonds denote a global shear mode~\cite{overvelde2012compaction}.
%the simultaneous buckling of line modes in the cases of elastic and plastic buckling triggers a single-step global buckling as the increase of line-mode number and ligament thickness, in contrast, a robust sequential buckling of line modes shows in a large range of line modes and ligament thickness yield buckling.
}
\label{fig:3}
\end{figure*} 

We apply the theory of elastic buckling and plastic buckling~\cite{cedolin2010stability,cimetiere2019coupling,shanley1947inelastic} to our unit cell and we find that yield buckling occurs in the regime defined by (see Appendix A):
\begin{equation}
\frac{4E_t/E}{\left(1+\sqrt{E_t/E}\right)^2}\frac{t}{2\ell}    <\frac{\sigma_y}{E}< \frac{t}{2\ell}
\label{eq:snapthroughprediction}
\end{equation}
To verify this prediction, we conduct numerical simulations by varying the parameters, $E_t/E$ and $t/\ell$ (Fig.~\ref{fig:2}C, see also varying $\sigma_{y}/E$ and $t/\ell$ in the Fig.~\ref{fig:8}), we find a good agreement by measuring the buckling stiffness and the post-buckling load before self-contact (red triangles in Fig.~\ref{fig:2}C and ~\ref{fig:8}C). From this analysis, it appears that when Eq.~\eqref{eq:snapthroughprediction} holds, the ligament plastifies at a load that is higher than the buckling load defined by the reduced modulus ~\cite{cedolin2010stability}---the left-hand side of Eq.~\eqref{eq:snapthroughprediction}. Since the plastified ligament has a much lower stiffness $E_t\ll E$, the unit cell will immediately lose its stability and buckle precisely at the onset of plasticity. In turn, the stress distribution across the ligament becomes highly asymmetric---the compressive part of the ligament is undergoing plastic loading, whereas the tensile part of the ligament is undergoing elastic unloading (Fig.~\ref{fig:9}F and Supplementary Movie 1). In the limit where the tangent modulus is much smaller than the elastic modulus, the elastic unloading dominates the total load and results in a load decrease at buckling, hence triggering yield buckling.

This load decrease will continue deep in the postbuckling region when the unit cell is further compressed until the two squares enter into contact (Fig.~\ref{fig:2}B inset). At this point, the unit cell dramatically stiffens. 
The conjunction of yield buckling and stiffening at contact is the key to achieving an orderly buckling in sequence when multiple unit cells are connected in series (Fig.~\ref{fig:2}D and Supplementary Movie 1).   
%We show in the SI that only yield buckling
%The decrease in load right at the point of buckling enables the design of sequential buckling events in metamaterials that host multiple buckling modes in clusters of unit cells. 
The load decrease ensures that the buckling of subsequent unit cells is not activated before the first buckled unit cells have made self-contact (Fig.~\ref{fig:2}EF). This cleanly delineated sequence is in contrast with the mixed buckling modes that occur in the case of elastic buckling or plastic buckling (Fig.~\ref{fig:9}JKL).

\begin{figure*}[t!]
\centering
\includegraphics[width=1.0\linewidth]{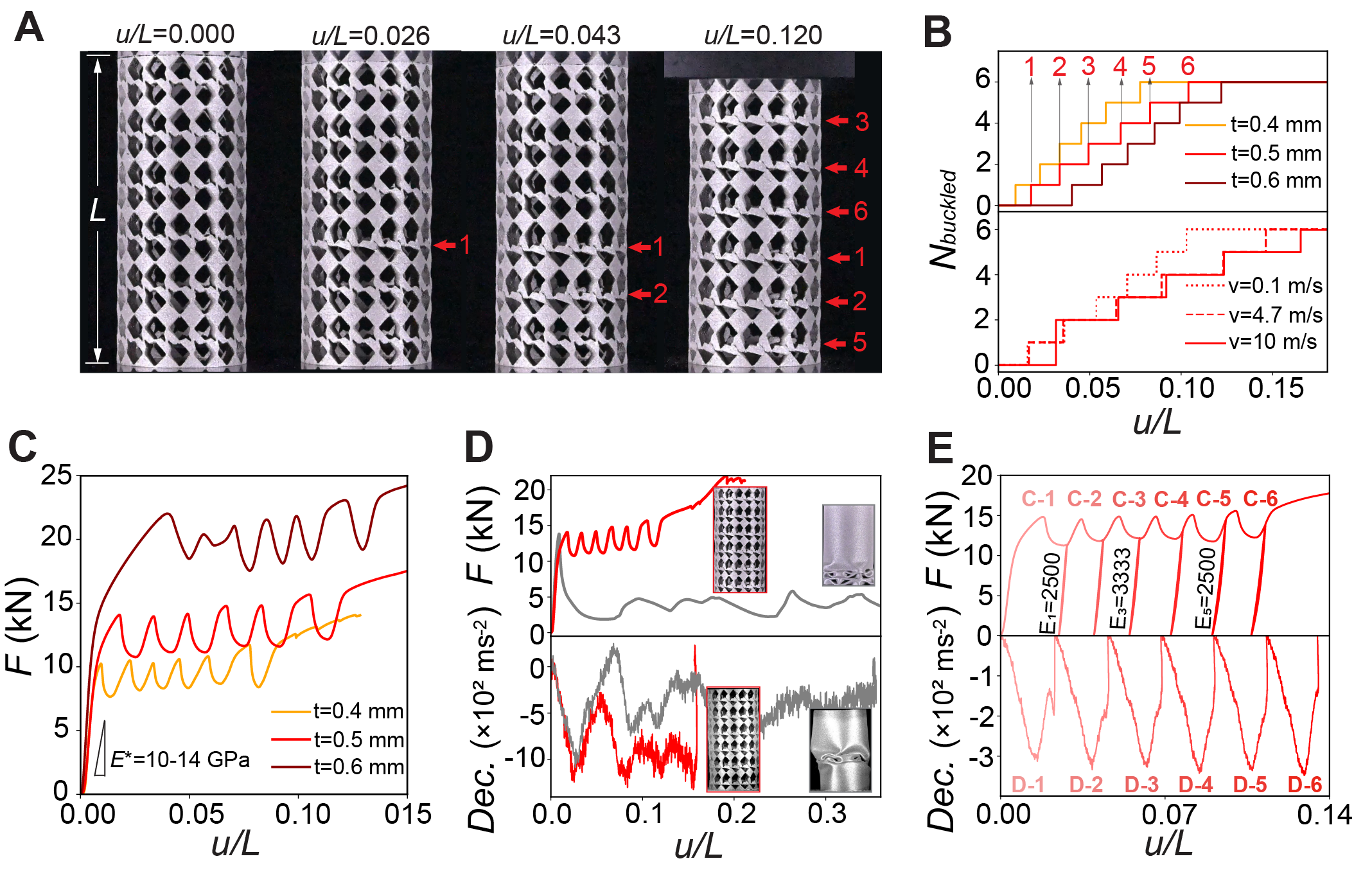}
\caption{\textbf{Experimental demonstration of ideal shock absorbers.} The ligament thickness of the metacylinder is $t=0.5$ mm and the compression speed is 0.2 mm/s unless indicated otherwise.
\textbf{A.} 
%a metamaterial with six-layer modes is configured into a cylinder shape as a shock absorber and 3D printed with 316L stainless steel, which shows six steps of sequential layer modes in the case of yield buckling.
Snapshots of the metacylinder under uniaxial compression at different strokes $u/L$.
\textbf{B.} 
%this sequential buckling is not only robust in a range of ligament thickness but a large range of loading speed from $0.1\; m/s$ to $10\; m/s$. 
%
Number of buckled line modes $N_\textrm{buckled}$ vs. compressive stroke $u/L$ for various ligament thicknesses (top) and loading speeds (bottom).
\textbf{C.} 
%the six steps of sequential buckling contribute to a stable plateau consisting of six numbers of small peaks in force-strain curves of the shock absorbers where the level of the plateau can be tuned by the ligament thickness. 
Force $F$ vs. compressive stroke $u/L$ for metacylinders with various ligament thicknesses.
\textbf{D.} Force vs. compressive stroke $u/L$ under quasi-static compression (top) and deceleration vs. compressive stroke under impact (bottom) for the metacylinder (red) and crash can (gray). 
%with similar stiffness and strength where the static compressing speed is $0.2\; mm/s$ and 
The impact speed is $4.7\; m/s$ with a weight of $15.5\; kg$. 
%The metacylinder shock absorber can prove stable shock absorption at the beginning of the impact and can absorb energy efficiently.
%
\textbf{E.} 
Force vs. compressive stroke displacement curve under six quasi-static compression cycles $C$ of increasing magnitude (top) and deceleration vs. compressive stroke under six distinct dynamic drops $D$ (bottom). See also Supplementary Movie 3.
%multi-usability of the shock absorber is proved by the static loading-unloading test and dynamic multiple dropping tests with a dropping speed, $3.2\; m/s$, and dropping weight, $5.5\; kg$., where similar stiffness and strength show in each loading process.
}
\label{fig:4}
\end{figure*} 

So far, we have shown that yield buckling enables sequential buckling with an arbitrary large number of steps. Next, we need to pair this concept with a metamaterial geometry that is structurally stable and that turns this sequential buckling into a force plateau of large magnitude.
%So far, we have achieved a stiff structure that exhibits sequential buckling with an arbitrary large number of steps.
%in the multi-layer rotating squares is only possible with yield buckling, providing repeated buckling behaviors and identical mechanical properties in each step. 
%Nevertheless, we remain far 
%falls short of 
%from an ideal shock absorber for two main reasons.
%First, our design requires an additional support structure to constrain global buckling (Fig.~\ref{fig:2}D). Second, the force curve does not exhibit the sought-after plateau and instead significantly drops after each buckling step (Fig.~\ref{fig:2}F).
To this end, we introduce a metamaterial geometry capable of preventing global shear buckling and of exhibiting many local buckling modes in series with limited force drops (Fig.~\ref{fig:3}A). The geometry is a variant of a flexible metamaterial geometry well known for avoiding global shear modes~\cite{overvelde2012compaction,overvelde2014relating} and that contains in addition modes localized along lines---line modes~\cite{bossart2021oligomodal,van2022machine,van2023emergent}. As in the case of the unit cell studied above, the same three regimes of buckling occur based on the aspect ratio $t/\ell$ and moduli ratio $E_t/E$. In contrast with the above unit cell, additional stabilizing ligaments---not in the loading path---tend to delay the onset of buckling.
%a to stretch when the unit cell buckles. 
%These stabilizing ligaments help to delay the buckling load beyond the reduced modulus load, making the structure buckle together with a negative stiffness beyond the yield buckling region
As such, the left inequality of Eq.~\eqref{eq:snapthroughprediction} is satisfied even if the hinge plastifies before buckling. Therefore the load decreases at buckling (Fig.~\ref{fig:3}B). However, as the stabilizing 
ligament is bending and stretching after buckling, this decrease is limited and the load even tends to increase further into postbuckling. Yield buckling---viz. the load remains below the initial buckling load at the point of self-contact---only occurs if the moduli ratio $E_t/E$ is sufficiently small (Fig.~\ref{fig:3}C red triangles). 
%To obtain good shock absorption properties in combination with high structural stability, it is desirable to limit the decrease in load prior to self-contact without adversely affecting the stroke of the shock absorption. 

%exhibiting a strip-like buckling mode (line mode) under compression(Fig.~\ref{fig:3} A). By incorporating extra ligaments in the horizontal directions, we expand the area of buckling with negative stiffness\cite{hutchinson1974plastic} and prevent the large force drop after buckling(Fig.~\ref{fig:3}B).  To validate this approach, we conduct a series of simulations with varying ratios of $E_{t}/E$ and $t/\ell$. We measure the stiffness and force values at the buckling point and the post-buckling point before the self-contact (Fig.~\ref{fig:3}B). Our findings indicate that introducing plasticity($E_{t}/E<1$) into the line-mode unit (Fig.~\ref{fig:3}C red) readily achieves the negative buckling stiffness of yield buckling. Moreover, achieving a post-buckling load before self-contact that is lower than the initial buckling load is only possible with a low tangent modulus ($E_{t}/E \ll 1$, Fig.~\ref{fig:3}C red triangles).

When two unit cells are connected in series, we observe that indeed 
the clean sequence of buckling and contact-induced stiffening is %only observed in the regime of yield buckling (Fig.~\ref{fig:3}DEF), whereas buckling oc
%We further demonstrate that achieving robust sequential buckling behavior in metamaterials 
only possible in the regime of yield buckling. 
%To prove this, we first examine a two-layer meta-structure with two line modes. 
In the case of elastic buckling with a positive buckling slope, both line-mode layers buckle simultaneously, followed by a force increase (Fig.~\ref{fig:3}D and Supplementary Movie 2). For plastic buckling, the negative buckling slope allows one line mode to buckle first, triggering the other line mode before self-contact as the post-buckling load quickly exceeds the initial buckling load (Fig.~\ref{fig:3}E and Supplementary Movie 2). Only in the case of yield buckling, with its drop in load until self-contact, can these two line modes be fully separated and buckle in a two-step sequence. Importantly, the load decrease is limited by the stabilizing ligament (Fig.~\ref{fig:3}F and Supplementary Movie 2). 
%Thereby the metamaterial geometry allows 
%To obtain good shock absorption properties in combination with high structural stability, it is desirable 
%to limit the decrease in load prior to self-contact without adversely affecting the stroke of the shock absorption. 

%We know show that
Surprisingly, %this metamaterial geometry in combination with 
yield buckling also allows to avoid unwanted global buckling modes of larger metamaterials and enables robust sequential line-mode buckling of an arbitrary large number of steps (Fig.~\ref{fig:10} and  Supplementary Movie 2), with a moderate decrease of load between the steps without adversely affecting the stroke of the shock absorption. Hence this precise combination is a particularly promising prospect for creating ideal shock absorbers.
We then experimentally prove that such metamaterials enabled by yield buckling behave as ideal shock absorbers. % in a robust, efficient, and reusable way.
We shape such 2D metamaterial pattern with six-layer modes into a cylinder. To trigger the yield buckling of the layer modes under uniaxial compression, we 3D print such metacylinder (Fig.~\ref{fig:4}A) with a steel that has a low tangent modulus $E_t=500$~MPa with respect to its Young's modulus $E=200$~GPa (316L stainless steel see Fig.~\ref{fig:12} for a calibration). 
During the compression, all the layers of the metacylinder initially deform. The deformation is
localized in the vertical ligaments, which leads to a high stiffness $E^{*}=10-14$~GPa (Fig.~\ref{fig:18}B). 
%(around 10 GPa, Fig.~\ref{fig:4} A green curve). 
Subsequently, at a compressive stroke 
$u/L=0.017$ one layer starts to buckle, and the rigid elements making up the layer start to counter-rotate, consistently with the simulation shown in Fig.~\ref{fig:3}A. 
%which corresponds to the first local maximum in force (red curve in Fig.~\ref{fig:4}B).
For further compression, the elements further rotate
%, and the force decreases 
until they enter into contact at $u/L=0.026$  (Fig.~\ref{fig:4}A).
%of the rotating squares is achieved, reaching a local minimum. 
The metacylinder thereby stiffens and for $u/L=0.033$, another layer buckles. The same process repeats as the metacylinder is further compressed until all six layers are fully collapsed at $u/L=0.12$ (red curve in Fig.~\ref{fig:4}B top). 
For further compression still $u/L=0.27$ the metacylinder fractures and buckles in a global buckling mode (Supplementary Movie 3). Hence in the range of strokes $0<u/L<0.12$, the metacylinder exhibits a delineated sequence of steps as desired. 

Each buckling step corresponds to a local maximum in the force curve (red curve in Fig.~\ref{fig:4}C) and each self-contact event to a local minimum. This sequence of six pairs of local maxima and minima makes up a wiggly plateau. Altogether this plateau and the high initial slope approximate very well the ideal shock absorber we have been after all along (Fig.~\ref{fig:1}A top-right).
%producing the plateau consisting of a series of local maxima and minima (Fig.~\ref{fig:4}B). 
%

Interestingly, this sequential shock-absorbing deformation is robust under various conditions.  The multistep sequential behavior remains robust over a wide range of different ligament thicknesses (Fig.~\ref{fig:4}B top), loading speeds (Fig.~\ref{fig:4}B bottom) and under off-axis compression (Fig.~\ref{fig:15}). 
Importantly, such ideal shock-absorbing behavior is tunable. We can adjust the shock absorption performances by simply changing the thickness of the ligament $t$ without significantly increasing the mass of the metacylinder (Fig.~\ref{fig:4}C and ~\ref{fig:18}).

\begin{figure}[t!]
\centering
\includegraphics[width=1.0\linewidth]{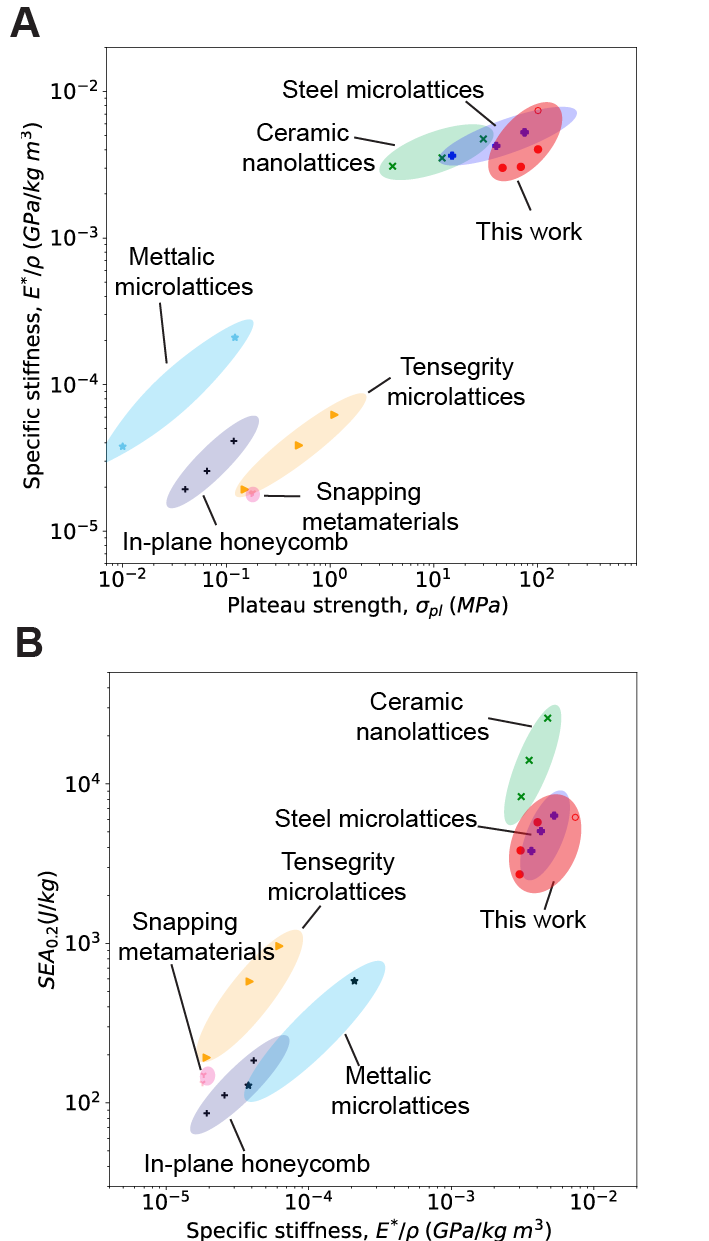}
\caption{\textbf{Ashby maps}
(\textbf{A}) Specific stiffness, $E^{*}/\rho$ vs. plateau strength, $\sigma_{pl}$
 and (\textbf{B}) specific energy absorption at $20\%$ strain $SEA_{0.2}$, vs. specific stiffness, $E^{*}/\rho$ vs for our metacylinder (red disks) and optimised metamaterial (FEM only, red circle) and for existing metamaterials from references~\cite{tancogne2016additively, meza2014strong, schaedler2011ultralight, bauer2021tensegrity, rafsanjani2015snapping, papka1998experiments}.} 
%
%shock absorbing performance of line-mode metamaterials compared with other common cellular materials made of snap-through elements and honeycomb lattices. Our metamaterials show the ability to combine high shock absorption and high stiffness.
\label{fig:5}
\end{figure}

Furthermore, the progressive collapse of the metacylinder begets an efficient shock-absorbing mechanism. We benchmark it with a crash can---a standard shock-absorbing structure used in e.g. vehicles---that have the same stiffness and strength. In both the static compression and dynamic drop tests, the progressive collapse of the metacylinder leads to a load---equivalently deceleration---plateau that evenly distributes the energy absorption over the entire stroke (Fig.~\ref{fig:4}D). In contrast, the catastrophic collapse of the crash can leads to a significant drop in the load---equivalently deceleration---that concentrates most of the energy absorption near the onset of buckling. As a result, the crash can requires more than twice the stroke to absorb the same amount of energy.% as the metacylinder. 

Lastly, we demonstrate that our metacylinder can be used multiple times while retaining its initial mechanical properties. The idea is the following: during a first shock, some layers collapse to absorb the energy, but the structure retains its stiffness and strength and can absorb subsequent shocks if it has layers left. To prove this, we first perform six cyclic compressions of increasing displacement, buckling one additional layer at every cycle (Fig.~\ref{fig:4}E top). The metacylinder exhibits similar strength and stiffness at every cycle. 
Additionally, we drop a mass $M=5.5\; kg$ at an impact speed $3.2 \; m/s$ six times and observe the same deceleration in each cycle (Fig.~\ref{fig:4}E bottom).
%demonstrate that metacylinder can be absorb six times with the same deceleration in each drop in a speed of impact, $3.2 \; m/s$, and a mass of dropping, $M=5.5\; kg$  (Fig.~\ref{fig:4}E bottom). 
In contrast, the crash can loses its initial properties after the first drop, resulting in much lower decelerations and much larger strokes in each subsequent drop (Fig.~\ref{fig:14}BEF). Therefore, yield buckling enables ideal shock absorption in a tunable, robust, short-stroke, and reusable way.
But how do these performances compare to state-of-the-art metamaterials? We define metrics for ideal shock absorbers using Ashby charts: the specific stiffness $E^{*}/\rho$ vs. the plateau strength $\sigma_p$---viz. a good proxy for the deceleration---and the specific energy absorption at $20\%$ strain $SEA_{0.2}$ vs. specific stiffness $E^{*}/\rho$ (Fig.~\ref{fig:5}AB). Such metrics characterize the ability of materials to combine high stiffness and efficient deceleration (Fig.~\ref{fig:5}A) and high stiffness and high energy dissipation (Fig.~\ref{fig:5}B). We find that in these metrics, our shock absorbers outperform other proposed geometries for shock absorption, such as reusable metamaterials made of snap-through elements~\cite{restrepo2015phase, shan2015multistable, frenzel2016tailored, chen2021reusable} or honeycomb lattices~\cite{papka1998experiments, khan2012experimental}. The performance of our metamaterials are comparable to that of ceramic nanolattices~\cite{meza2014strong} and of metallic microlattices~\cite{tancogne2016additively}. 
Yet unlike these lattices, our metamaterials could readily be produced over a wide range of scale with multiple manufacturing methods and are inherently reusable multiple times while keeping their initial stiffness intact. %Unlike the metallic microlattices, our metamaterials could be made lightweight in a subsequent optimization step. 

% and could readily be mass-manufactured.
%We demonstrate that the sequential yield buckling provides superior impact absorption performances: the metacylinders can combine high stiffness and high energy absorption and can also be reused multiple times whilst retaining their initial stiffness and strength.

%So far, we have proved that the metamaterials enabled by yield buckling perform as ideal shock absorbers, yet only along one direction. We conclude this article by designing orthogonal layers in two or three dimensions (Fig.~\ref{fig:5}C-F), which allow for enhanced shock absorption in two or three directions.
%buckling into two directions by using a 2D unit with two line modes. (Fig.~\ref{fig:5}A). 
%In two dimensions, we modify the design of Fig.~\ref{fig:3}A to host line modes along two directions. Our simulations confirm that the layers buckle in sequence when compressed in both directions respectively (Fig.~\ref{fig:5}D and Supplementary Video XXX). As a result, the force-displacement curves combine high stiffness prior to buckling and a wiggly plateau with 6 oscillations that induce high dissipation (Fig.~\ref{fig:5}E).
%We then approve it with the simulation in a 2D metamaterial with $6 \times 6$ orthogonal line modes where the same 6 localized line modes behavior shows in X and Y directions (Fig.~\ref{fig:4}B). 
%We generalize the concept to a 3D unit cell with three orthogonal layer modes (Fig.~\ref{fig:5}F) and perform a mode analysis that confirms the existence of orthogonal layers in all three directions (Fig.~\ref{fig:5}G).
             
%\section{Outlook}

In conclusion, we have demonstrated that yield buckling, in combination with suitable metamaterial architectures, is a crucial tool for achieving sequential buckling and creating ideal shock absorbers. 
Importantly our ideal shock absorbers can absorb more impact energy in a smaller volume, can be made from any elastoplastic material, can be generalized to impacts in multiple directions (see Appendix C for demonstration), and could in principle be mass-manufactured (see Appendix B). Our ideal shock absorbers could hence be applied in a range of applications, from automotive and aerospace at the meter size to microscopy and nano-lithography at the micrometer size, where there are strong drivers for safety and sustainability. Exciting questions ahead are how to tailor them to specific applications and optimize them for either higher performances. More generally, our work suggests that using material nonlinearity in addition to geometry considerably enriches the toolbox of metamaterials and will ultimately allow them to pervade into real-world applications.

%Exciting research questions ahead are to optimize our metamaterials for even higher performances, to manufacture them at the large scale, to expand our concept beyond stainless steel to a wide range of materials, and to tailor them to specific high-tech applications.

%\cc{add comment bernard. come back on the more efficient impact absorption (less buckling stroke). on the fact that these can be applied to any elastoplastic material and to the fact that they can be made without 3d printing so could be mass-manufactured.}
%demonstrated that the plasticity of materials combined with the buckling modes of geometry can be used to design ideal shock absorbers with a multi-step sequential deformation. 

%opens a new avenue for 
%using plasticity as a design tool and for the application of metamaterials in shock damping.

\emph{Acknowledgments.} 
We thank Eisso Atzema, Israel Pons, Sebastien Neukirch and Corrado Maurini for insightful discussions and suggestions, Daan Giesen, Clint Ederveen Janssen and Jan Heine for technical assistance. We acknowledge funding from the European Research Council under grant agreement 852587 and the Netherlands Organisation for Scientific Research under grant agreement NWO TTW 17883.

%All the codes and data supporting this study are available on the public repository  \url{https://doi.org/10.5281/zenodoXXXXX}.
% \nolinenumbers
%\clearpage
\bibliographystyle{naturemag}

%\bibliography{Reference}

\clearpage
\setcounter{equation}{0}
\renewcommand{\theequation}{S\arabic{equation}}%
\setcounter{figure}{0}
\renewcommand{\thefigure}{S\arabic{figure}}%
\renewcommand{\figurename}{\EDF}%

\onecolumngrid
\begin{appendix}

In this appendix, we show how to exploit yield buckling as a design concept for metamaterials with sequential buckling (Appendix A). Then in Appendix B, we present geometrical design, fabrication, finite element simulations protocols and experimental protocols. Finally, in Appendix C, we present additional complementary numerical and experimental data.

\section{Yield buckling as a metamaterial concept}
In this appendix section, we introduce the notion of plastic buckling and introduce the concept of yield buckling theoretically and numerically. We then investigate how yield buckling enables sequential buckling in multiple steps.

\subsection{Buckling behaviors with elastoplastic material}
The buckling behavior of a single column has been studied for more than 250 years. Euler first studied the elastic buckling behavior of a single beam and found the elastic bifurcation load is proportional to the elastic modulus and inversely proportional to the slenderness of the column~\cite{euler1952methodus}. When the column is not very slender, it can fail not only due to instability but also to a combination of instability and material failure like plastic yielding when the material is elastoplastic. Considere and Van Karman first studied this buckling behavior and calculated the instability load in the plastic loading process, known as the reduced modulus load $F_r$~\cite{shanley1947inelastic, cedolin2010stability, hutchinson1974plastic}. The reduced modulus load $F_r$ is the upper bound for the plastic buckling load and assumes that the loading and unloading stresses in either side of the beam's cross section are equal and opposite. The loading part of the beam's cross-section undergoes plastic loading, while the unloading part of the beam's cross section undergoes elastic unloading.
%based on the elastic unloading part in the cross-section being equal to the plastic loading part. 
37 years later, Shanley~\cite{shanley1947inelastic} introduced the tangent modulus load $F_t$. The tangent modulus load is a lower bound for the plastic buckling load. In contrast to $F_r$, the hypothesis is that the whole area of the cross-section in the column undergoes plastic loading. The tangent load $F_t$ is therefore always smaller than the reduced modulus load $F_r$ and whenever 
the yield stress $\sigma_y<F_t /A $, where $A$ is the cross section area of the beam. the column will undergo plastic buckling. In postbuckling the load typically increases until it reaches a maximum load $F_{max}<F_r$. %Deep in this far postbuckling regime, both unloading and loading parts of the cross-section undergo plastic deformations.

%In the postbuckling of tangent modulus theory,  the structure has a positive stiffness 
 %$F_{t}<F_{max}<F_{r}$. 
 %The maximum load $F_{max}$ in practice is usually closer to the tangent modulus load rather than the reduced modulus load \cite{shanley1947inelastic}. 
%Furthermore, Hutchinson analyzed the imperfection effect on the post-buckling based on Shanley's model and found that the imperfection can be magnified in the plastic buckling by the ratio, $E/E_{t}$, which can lower the buckling load from the plastic stage to elastic stage. However, the use of yield load in plastic buckling has not been well exploited. 

These two cases assume that the column plastifies before buckling. Yet, there are additional cases, where the onset of plasticity induces buckling instead. 
In a first case, the yield load is larger than the tangent modulus load $\sigma_y>F_t /A$, there is no solution for the tangent modulus load and the column will buckle at the yield point followed by a slight force increase~\cite{cedolin2010stability}. 
In a second case, the yield load is larger than the reduced modulus load $\sigma_y>F_r /A$~\cite{cedolin2010stability,cimetiere2019coupling}. In this regime, which we call yield buckling, the load will strictly decrease in postbuckling.

In conclusion, whenever the yield load is smaller than the reduced modulus load, $\sigma_y<F_r /A$, an elastoplastic column will undergo plastic buckling and exhibit a positive postbuckling stiffness. Whenever $\sigma_y>F_r /A$, an elastoplastic column will undergo yield buckling and exhibit a negative postbuckling stiffness. In the following, we apply this plastic buckling theory to the case of the metamaterials unit cell and derive Eq.~\eqref{eq:snapthroughprediction} of the Main Text.

%In this work, we study a regime of buckling load above the reduced modulus load which shows strictly negative stiffness right at the onset of buckling. We further demonstrate such a regime can be used to design ideal shock absorbers in metamaterials.

\subsection{Yield buckling with rotating-square unit cell}

\subsubsection{Three buckling regimes}

\begin{figure*}[t!]    
\centering
\includegraphics[width=0.3\linewidth]{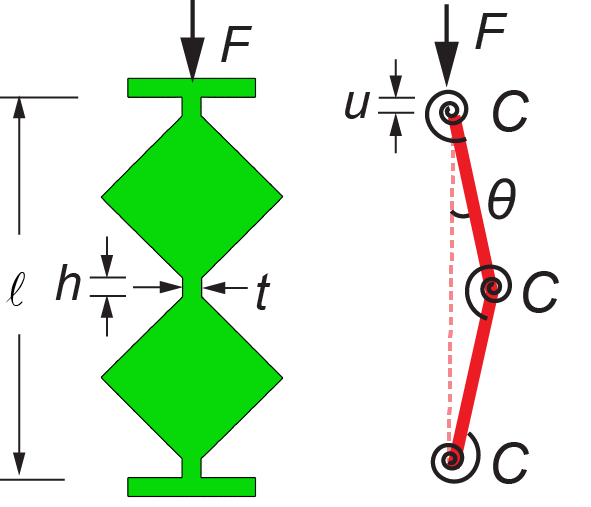}
\caption{\textbf{Sketch of buckling unit cell}. }
\label{fig:6}
\end{figure*}

In this section, we will analyze the buckling behavior of a rotating-square unit cell with elastoplastic material (Fig.~\ref{fig:2}A) and define the yield buckling regime. To simplify the analysis, we assume that the rotating squares are undeformable and the ligaments of the unit cell have the same dimension in both height and width directions, $h=t$ (Fig.~\ref{fig:6}). We identify three regimes.

\begin{figure*}[t!]    
\vspace{-3mm}
\centering
\includegraphics[width=1.0\linewidth]{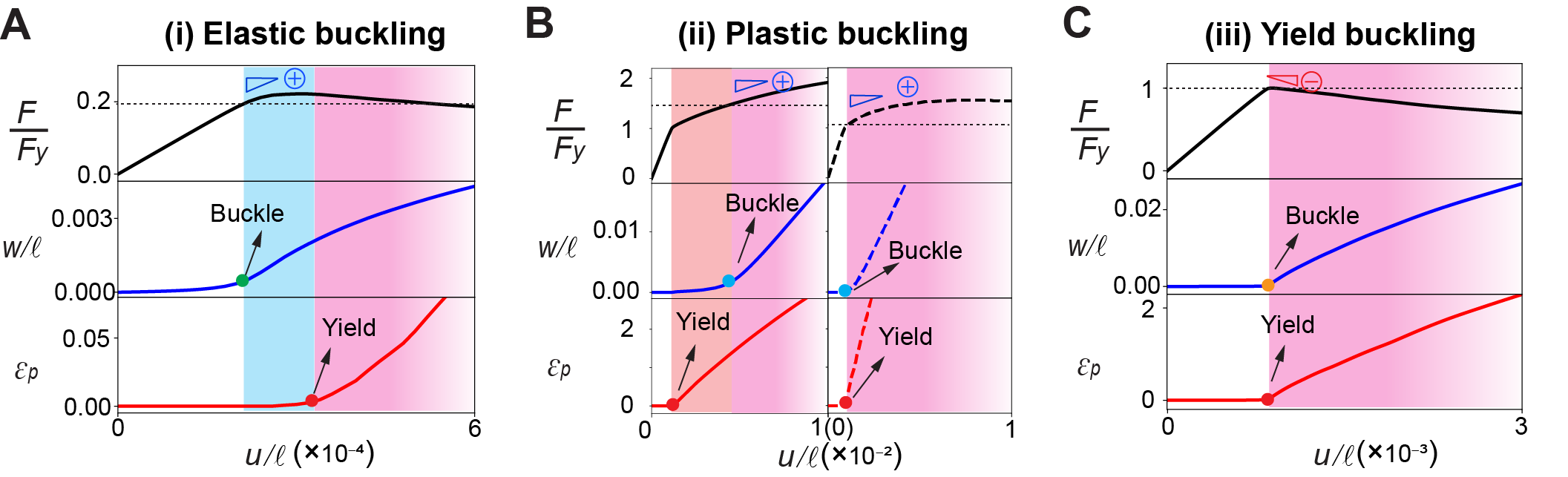}
\caption{\textbf{Buckling behaviors of the unit cell with elastoplastic material}. Normalized force $F/F_{y}$, is used to determine the stiffness at the buckling point (black curve), normalized deflection $w/\ell$ is used to determine the buckling point (blue curve), and the plastic strain $\varepsilon_{p}$ is used to determine the plasticity at the ligaments(red curve). \textbf{A,} elastic buckling, the sky-blue shaded area denotes buckling without plasticity and the magenta shaded area denotes buckling with plasticity. \textbf{B,} plastic buckling, red fill denotes plasticity without buckling. \textbf{C,} yield buckling. In (A) and (B), the stiffness at the onset is positive whereas it is negative in (C).}
\label{fig:7}
\end{figure*}

\paragraph{Regime (i): elastic buckling}
When the aspect ratio $t/\ell$ is small enough or the material model is purely elastic, the unit cell buckles before the ligaments plastify.
We then first analyze the elastic buckling behavior of this unit cell. 
The unit cell has the potential energy $\Pi$ under a force $F$:  
%\begin{equation}\label{potential energy}
$\Pi=C\theta ^{2} + \frac{1}{2} C(2\theta) ^{2} - Fu$
%\end{equation}
where $\theta$ is the rotating angle of the squares, $u$ is the compressing displacement. For a small angle $\theta$, $u=\ell(1-cos(\theta)) \approx \ell\theta ^{2}/2$. From the equilibrium condition, $\partial\Pi/\partial\theta=0 $, we can obtain the elastic buckling load,
%\begin{equation}\label{elastic buckling load}
$F_{e}=\frac{6C}{\ell}=\frac{Et^{2}}{2\ell}$.
%\end{equation}
The post-buckling behavior of the buckling unit cell can be obtained from force balance,
%\begin{equation}\label{post elastic buckling load}
$F=\frac{\theta}{\sin \theta} F_{e}$
%\end{equation}
where the unit cell shows a small positive stiffness after buckling since $(\theta/\sin\theta) \approx 1/(1-\theta^{2}/2)>1$ (see simulations in Fig.~\ref{fig:2}B (green) with a linear elastic material model and Fig.~\ref{fig:7}A with an elastoplastic model).
For the elastoplastic material model, the yield load of the unit cell only depends on the yield stress of the material $\sigma_{y}$ and the cross-section area of the ligament $A=t\times1$.
Here, we can use the yield load $F_{y}=\sigma_{y}t$ and elastic buckling load $F_{e}$ to define the regime of elastic buckling (regime (i)) as the buckling happens before the ligaments plastify % and has a positive stiffness at the onset of buckling where
$F_{e}<F_{y}$, or equivalently 
\begin{equation}\label{regime I}
\frac{\sigma_{y}}{E}>\frac{t}{2\ell}. 
\end{equation}
The stiffness is positive at the onset of buckling~\cite{cedolin2010stability,cimetiere2019coupling}.

\paragraph{Regime (ii): plastic buckling}
In this regime, there are two sub-cases. 

First, when the aspect ratio $t/\ell$ is large, the unit cell can buckle after the ligaments plastify and buckling starts at the critical buckling load, yet where the elastic modulus $E$ is replaced by the tangent modulus $E_{t}$, which is called the tangent modulus load:
%\begin{equation}\label{tangent modulus load}
$F_{t}=\frac{E_{t}t^{2}}{2\ell}$.
%\end{equation}
The tangent modulus load provides a lower bound for plastic buckling. The postbuckling stiffness of the unit cell is positive (Fig.~\ref{fig:7}B solid curves). Hence, this sub-case occurs in the range $F_{y}<F_{t}$, or equivalently
%Here we can define the regime of plastic buckling (regime (ii)) as the regime where the unit cell buckles followed by a positive stiffness (Fig.~\ref{fig:6}B):
\begin{equation}\label{regime II-1}
\frac{\sigma_{y}}{E}< \frac{E_t}{E}\frac{t}{2\ell}.
\end{equation}
%where $\xi=E_{t}/E$. 

Second, 
%
%The tangent modulus load is only valid when the unit cell buckles after the ligament plastify.
when the yield load is larger than the tangent modulus load, $F_{t}<F_{y}$, since the column has not plastified yet, $F_{t}$ is not a valid buckling load. Instead, the unit cell will buckle at the yield load followed by a positive stiffness after buckling (Fig.~\ref{fig:7}B dashed curves). 
The upper bound for plastic buckling is given by the critical buckling load, yet where the elastic modulus $E$ is replaced by the reduced modulus $E_{r}$, which is called the reduced modulus load: 
%\begin{equation}\label{reduced modulus load}
$F_{r}=\frac{E_{r}t^{2}}{2\ell}$, 
%\end{equation}
where $ E_{r}=4EE_{t} / (\sqrt{E_{t}}+\sqrt{E})^{2}$ \cite{cedolin2010stability}. Hence, this sub-case occurs in the range $F_{t}<F_{y}<F_r$, or equivalently,
\begin{equation}\label{regime II-2}
 \frac{E_t}{E} \frac{t}{2\ell}<\frac{\sigma_{y}}{E}<\frac{4E_t/E}{\left(1+\sqrt{E_t/E}\right)^2}\frac{t}{2\ell}
\end{equation}
For both cases of buckling in the elastic and plastic regimes (regimes (i) and (ii)), the unit cell exhibits a positive stiffness at the onset of buckling~\cite{cedolin2010stability,cimetiere2019coupling}.

\paragraph{Regime (iii): yield buckling}
When the yield load is larger than the reduced modulus load, $F_{r}<F_{y}$, since the column has not plastified yet, $F_{r}$ is not a valid buckling load. Instead, the unit cell will buckle at the yield load followed by a negative stiffness after buckling (Fig.~\ref{fig:7}C). The reason why the slope is negative is that $F_r$ is the upper bound for plastic buckling and the load can only decrease towards $F_r$.
Importantly, the postbuckling load always remains below the initial buckling load before the unit cells reaches self-contact. Hence, yield buckling occurs in the range $F_{r}<F_{y}<F_{e}$ or equivalently,
%Here, we can reach this regime by tuning the yield load in between the reduced modulus load and the elastic Euller load:
\begin{equation}\label{regime III-2}
\frac{4E_t/E}{\left(1+\sqrt{E_t/E}\right)^2}\frac{t}{2\ell}<\frac{\sigma_{y}}{E}<\frac{t}{2\ell}
\end{equation}
which is Eq.~\eqref{eq:snapthroughprediction} of the Main Text.
%We demonstrate this regime in the simulation (Fig.~\ref{fig:7}) where the unit cell buckles right at the onset of the ligaments plastify with a negative stiffness.

\subsubsection{Parametric study}

\begin{figure*}[t!]
\centering
\includegraphics[width=0.5\linewidth]{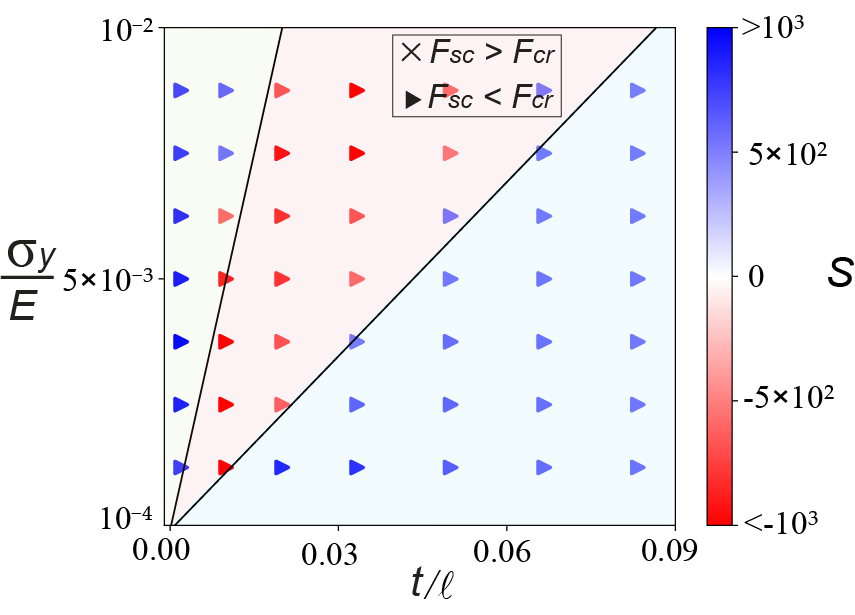}
\caption{\textbf{Post-buckling stiffness $S$ vs. aspect ratio of the unit cell $t/\ell$ and ratio between yield stress and Young's modulus $\sigma_{y}/E$.} 
The red triangles denote the yield buckling regime defined by $S<0$ and $F_{sc}<F_{cr}$. }
\label{fig:8}
\end{figure*}

We can distinguish the three regimes of buckling with two of the three dimensionless parameters, the aspect ratio $t/\ell$, the moduli ratio $E_{t}/E$, and the relative yield stress $\sigma_{y}/E$. In the Main Text, we have demonstrated that the three regimes of buckling can be reached by varying the aspect ratio $t/\ell$ and the moduli ratio $E_{t}/E$ with a given specific yield strength $\sigma_{y}/E=$ 0.5 GPa / 200 GPa (Fig.~\ref{fig:2}C). Here, the same concept can also be proven by tuning the relative yield stress $\sigma_{y}/E$ and the aspect ratio $t/\ell$ with a given value of moduli ratio $E_{t}/E$ = 20 GPa /200 GPa. By measuring the stiffness at the buckling and the postbuckling load before self-contact, we find a good agreement between the theory (green, blue, and red shaded areas correspond to elastic, plastic, and yield buckling regimes) and the simulation (triangles) (Fig.~\ref{fig:8}). The results confirm that only yield buckling (red triangles) allows the postbuckling load to always remain  below the initial buckling load before self-contact.

\subsubsection{Stress asymmetry}

\begin{figure*}[t!]     
\vspace{-2mm}
\centering
\includegraphics[width=0.96\linewidth]{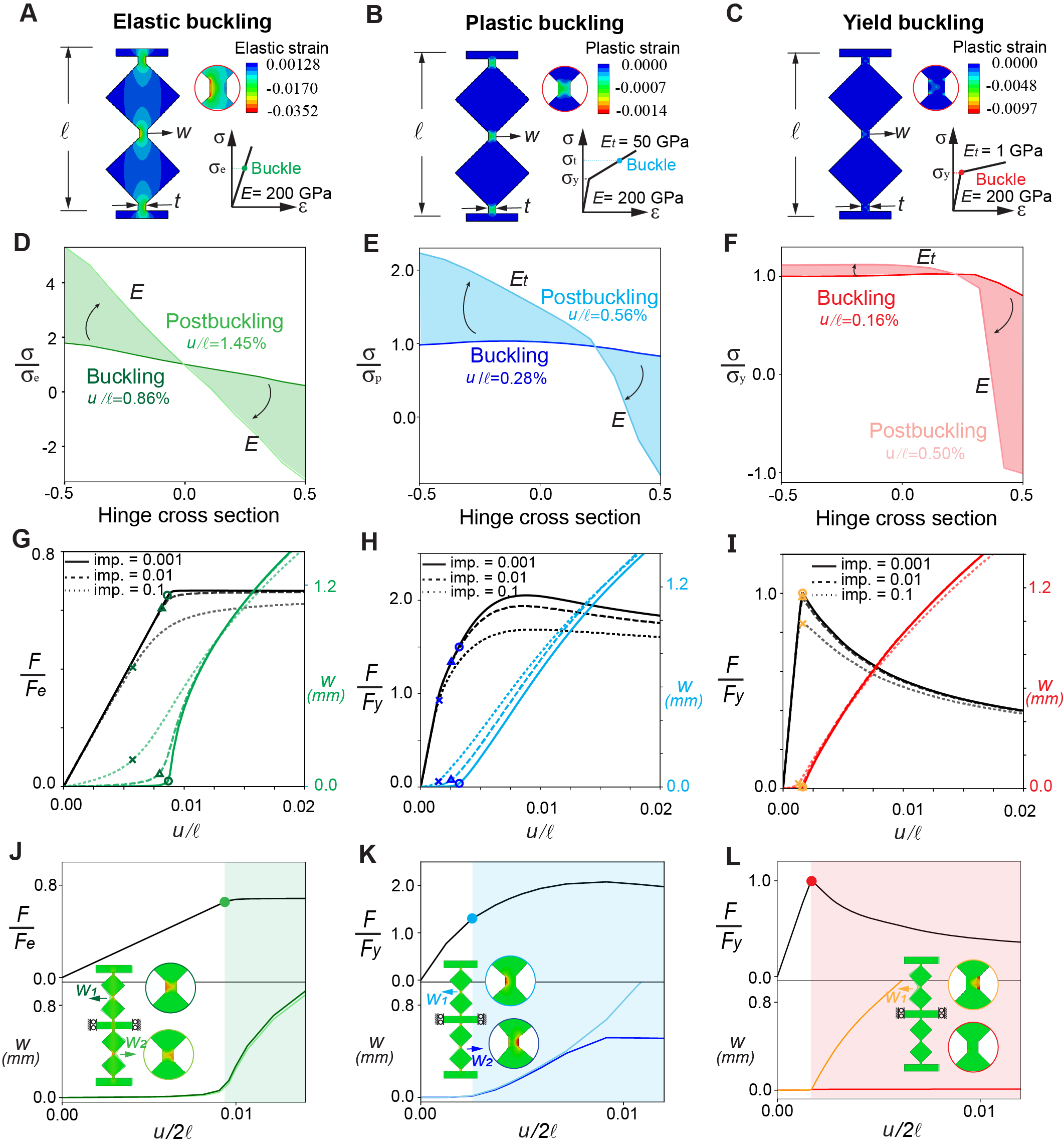}
\caption{\textbf{Yield buckling and imperfection study.} \textbf{A, B, C,} the strain contour at the moment of elastic, plastic, and yield buckling with the same geometry and different material models. \textbf{D, E, F,} the stress distribution along the section of the middle ligament from the buckling to a post-buckling moment.\textbf{G, H, I,} imperfection sensitivity analysis of the three regimes of buckling. Force (gray-scale lines) and lateral deflection (colored lines) vs. displacement $u/(2\ell)$. \textbf{J, K, L,} Force (top) and lateral deflection (bottom) vs. displacement $u/(2\ell)$. Inset: The buckling snapshot of the two lumped unit cells of the three regimes}
\label{fig:9}
\end{figure*}

To further explain the origin of the yield buckling, we analyze the strain and stress distributions across the ligament of the unit cell at the onset of buckling. The elastic and plastic strains at the elastic and plastic buckling points have a homogeneous distribution across all of the ligaments (Fig.~\ref{fig:9}AB). In contrast, the plastic strain contour at yield buckling shows that the plasticity at the onset of buckling is localized on the left side of the ligaments (Fig.~\ref{fig:9}C). This is because buckling and yielding are triggered simultaneously . The stress at the ligament in elastic buckling almost symmetrically distributes across the ligament during postbuckling (Fig.~\ref{fig:9}D) making the stiffness in the postbuckling slightly positive (Fig.~\ref{fig:9}G black solid curve). The stress distribution of the ligament's cross-section in plastic buckling and yield buckling both show an asymmetric distribution---the compressive part of the ligament is undergoing plastic loading with a slope, $E_{t}$, whereas the tensile part of the ligament is undergoing elastic unloading with a slope, $E$ (Fig.~\ref{fig:9}EF). A highly asymmetric stress distribution only appears in the yield buckling with moduli $E \gg E_{t}$, resulting in the load drops at the onset of buckling (Fig.~\ref{fig:9}I).

\subsubsection{Imperfection sensitivity} 

The imperfection sensitivity of the plastic buckling has been theoretically analyzed by Hutchinson \cite{hutchinson1974plastic} and the result shows a small imperfection in the plastic buckling can be magnified by the inverse of the moduli ratio $E_{t}/E$ compared with the elastic buckling.  Here, we conduct the imperfection sensitivity analysis with simulation for the yield buckling and the elastic and plastic buckling (Fig.~\ref{fig:9}GHI). Interestingly, yield buckling exhibits negative postbuckling stiffness even in the presence of imperfections.
%the yield buckling (Fig.~\ref{fig:9}I) with the largest ratio $E/E_{t}=200$ shows the smallest sensitivity to the geometry imperfection compared to the elastic buckling with $E/E_{t}=1$ (Fig.~\ref{fig:8}I) and plastic buckling with $E/E_{t}=4$ (Fig.~\ref{fig:9}H). 
%The buckling load of the yield buckling only reduces $15.7\%$ when the imperfection increases from $w_{0}=0.001$ to $w_{0}=0.1$. In contrast, the initial buckling load and the maximum postbuckling load of the plastic buckling are reduced by $39.1\%$ and $17.3\%$, respectively, (Fig.~\ref{fig:9}H) and these two values are $37.5\%$ and $6\%$ in the elastic buckling (Fig.~\ref{fig:9}G). ??? so what?

\subsubsection{Separating buckling events}

To further distinguish the difference in buckling mode separation between yield buckling and elastic and plastic buckling, we conduct simulations in a simple structure consisting of the two unit cells with the same geometry and material model as we analyzed above. And we find that the two unit cells buckle simultaneously with a force increase in the postbuckling in both elastic and plastic buckling (Fig.~\ref{fig:9}JK). In contrast, the buckling of the two unit cells fully separates from each other at the onset of buckling in the top unit cell (Fig.~\ref{fig:9}L). The buckling study of this two-layer structure proves that only yield buckling achieves clean sequential buckling, which can be used for designing a large number of sequential buckling (Fig.~\ref{fig:2}DEF).

\subsection{Global buckling vs. localized buckling}
\begin{figure*}[t!]
\vspace{-2mm}
\centering
\includegraphics[width=1.0\linewidth]{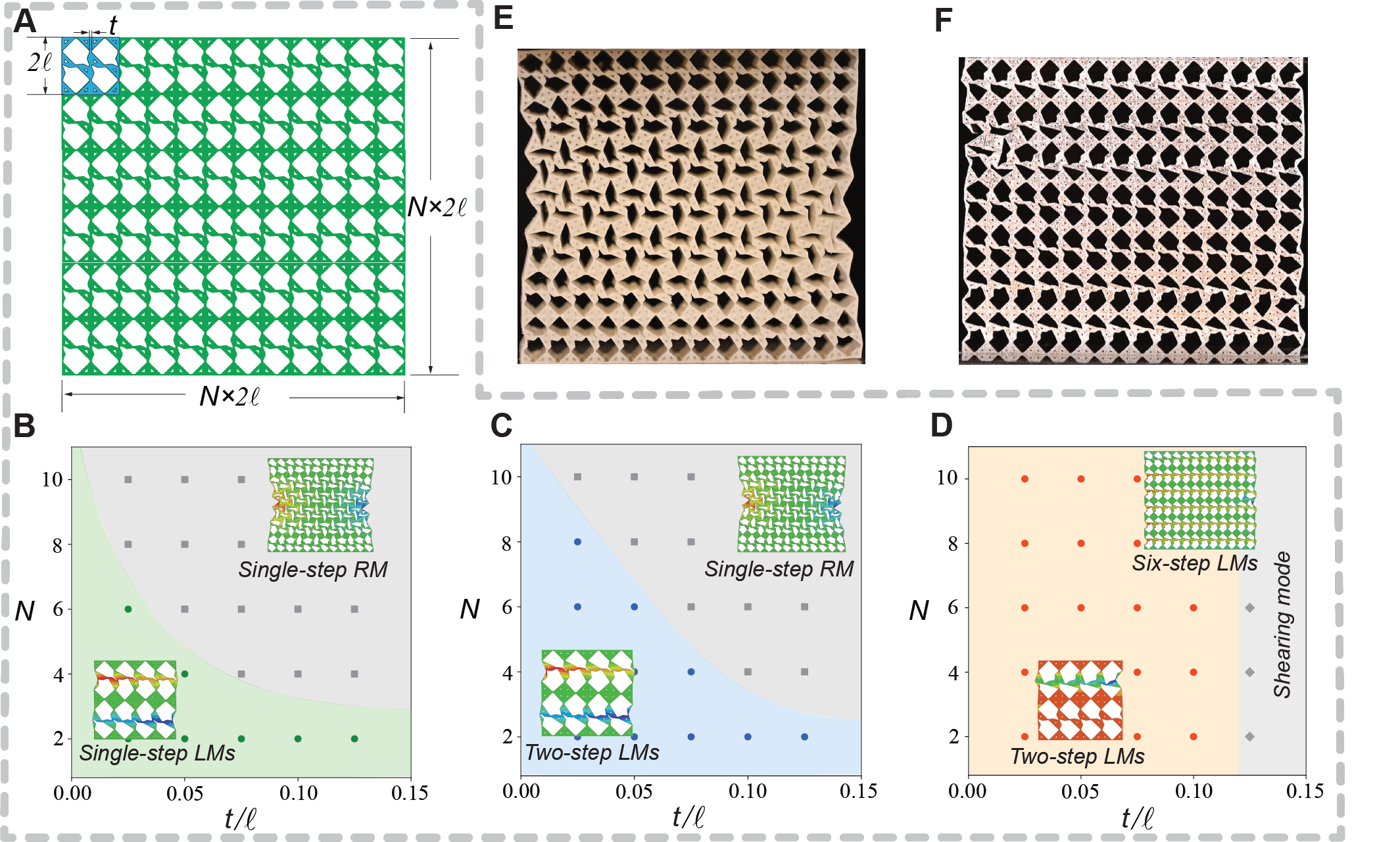}
\caption{\textbf{Yield buckling in large scale metamaterials.} \textbf{A,} A metamaterial consisting of $N \times N$ unit cells. \textbf{B,C, D,} Buckling mode vs. aspect ratio $t/\ell$ and metamaterials size $N$ in the elastic (\textbf{B}), plastic (\textbf{C}) and yield buckling (\textbf{D}) regimes from a finite element analysis. Grey squares denote a global counter rotating-square mode~\cite{resch1965geometrical,coulais2018characteristic,czajkowski2022conformal} in a single step (see insets panels \textbf{BC}), green and blue circles denote simultaneous buckling of all line modes (see insets in panels \textbf{BC}), red circles denote sequential buckling of the lines modes (see insets panels \textbf{D}) and grey diamonds denote a global shear mode~\cite{overvelde2012compaction}. \textbf{E,F} Snapshots of metamaterials with $7 \times 7$ unit cells under compression. (\textbf{B}) is 3D printed from elastic rubber and (\textbf{C}) is printed from 316L stainless steel.}
\label{fig:10}
\end{figure*}

We have demonstrated that yield buckling naturally leads to clean sequential buckling in multiplayer rotating-square unit cells constrained by the global shearing (Fig.~\ref{fig:2}DEF and ~\ref{fig:9}L). Here, we further show that yield buckling also allows to avoid unwanted global buckling modes in large-scale metamaterials. We assemble the line-mode unit cell (Fig.~\ref{fig:3}A) into a $N \times N$ metamaterial with a range of aspect ratios $t/\ell$ (Fig.~\ref{fig:10}A). 

In the regimes of elastic and plastic buckling, the local buckling modes are simultaneous. Crucially, those local buckling modes only occur in a regime where the number of unit cells in metamaterials is small and the ligament thickness is small (Fig.~\ref{fig:10}). Instead, for large metamaterials size $N$ and large ligament thickness, 
%For cases of simultaneous buckling (Fig.~\ref{fig:3} E) and sequential buckling without self-contact (Fig.~\ref{fig:3} D), 
the metamaterials show a global---rotating-squares---buckling mode 
%as the increase of line modes $N$ and the length ratio $t/\ell$ 
(Fig.~\ref{fig:10}BC and Supplementary Movie 2).
We experimentally illustrate such elastic global buckling in a 3D-printed rubber-like metamaterial with $7 \times 7$ numbers of unit cells (Fig.~\ref{fig:10}E).

In stark contrast, yield buckling enables sequential buckling of line modes for a wide range of sizes $N$ and ligament thickness  (Fig.~\ref{fig:10}D and Supplementary Movie 2). We experimentally illustrate such sequential buckling in a metamaterial made from 316L stainless steel under compression (Fig.~\ref{fig:10}F and see also the simulations in Supplementary Movie 2).
In conclusion, yield buckling and metamaterial geometry symbiotically promote robust sequential buckling of an arbitrary large number of steps.

\section{Simulations, Fabrication and Experiments}
In this appendix, we present the numerical protocols, metamaterials geometry and fabrication and experimental protocol. 

\subsection{Numerical simulation}
For the finite-element simulation of the buckling unit cell and the related metamaterials, we use the commercial software Abaqus(2020) with Standard model. 

\textit{Model definition}. Firstly, we model two-square unit cells and line-mode unit cells with an elastoplastic material model (Fig.~\ref{fig:2}A and Fig.~\ref{fig:3}A), varying the aspect ratio $t/\ell$ and the moduli ratio $E_{t}/E$ with a constant ligament thickness $t=1$ elastic modulus $E=200$ GPa, Poisson's ratio $\mu=0.3$, and yield stress $\sigma_{y}=500$ MPa (Fig.~\ref{fig:2}C). We also model the two-square unit cells, varying the aspect ratio $t/\ell$ and the relative yield strength $\sigma_{y}/E$ with a constant moduli ratio $E_{t}/E=0.1$ (Fig.~\ref{fig:8}). Secondly, we model multiple layers of two-square unit cells and metamaterials with line-mode unit cells, using $\sigma_{y}/E=0.0025$ and $E_{t}/E=0.0025$ for yield buckling. We use plane stress conditions with quadratic triangular elements (CPS6). We construct the mesh so that all the ligaments in the unit cells are 5 elements across.

\textit{Boundary condition}.
For two-square unit cells, line-mode unit cells, and line-mode metamaterials, we fix the bottom boundary and only allow vertical displacement at the top boundary. For multiple layers of two-square unit cells. we confine the side movement of the separated plates and only allow vertical movement.

\textit{Analysis}.
We perform two types of analysis: linear eigenmode analysis, where we calculate the lowest eigenmodes; Nonlinear bifurcation analysis, where a displacement imperfection from the linear analysis is introduced into the nonlinear compression (Static step). We make the imperfection proportional to the length of the unit cell $\ell$ and fix it to $0.001$ unless specified otherwise. We track the deflection $w$ of the middle ligament of the unit cell and use the bifurcation point on the deflection curve to catch the start of buckling. Then we measure the load, the plastic strain, and stress distribution across the ligament at the buckling point and the postbuckling process. 

\subsection{Geometry design and fabrication}
\begin{figure*}[t!]
\centering
\includegraphics[width=1.\linewidth]{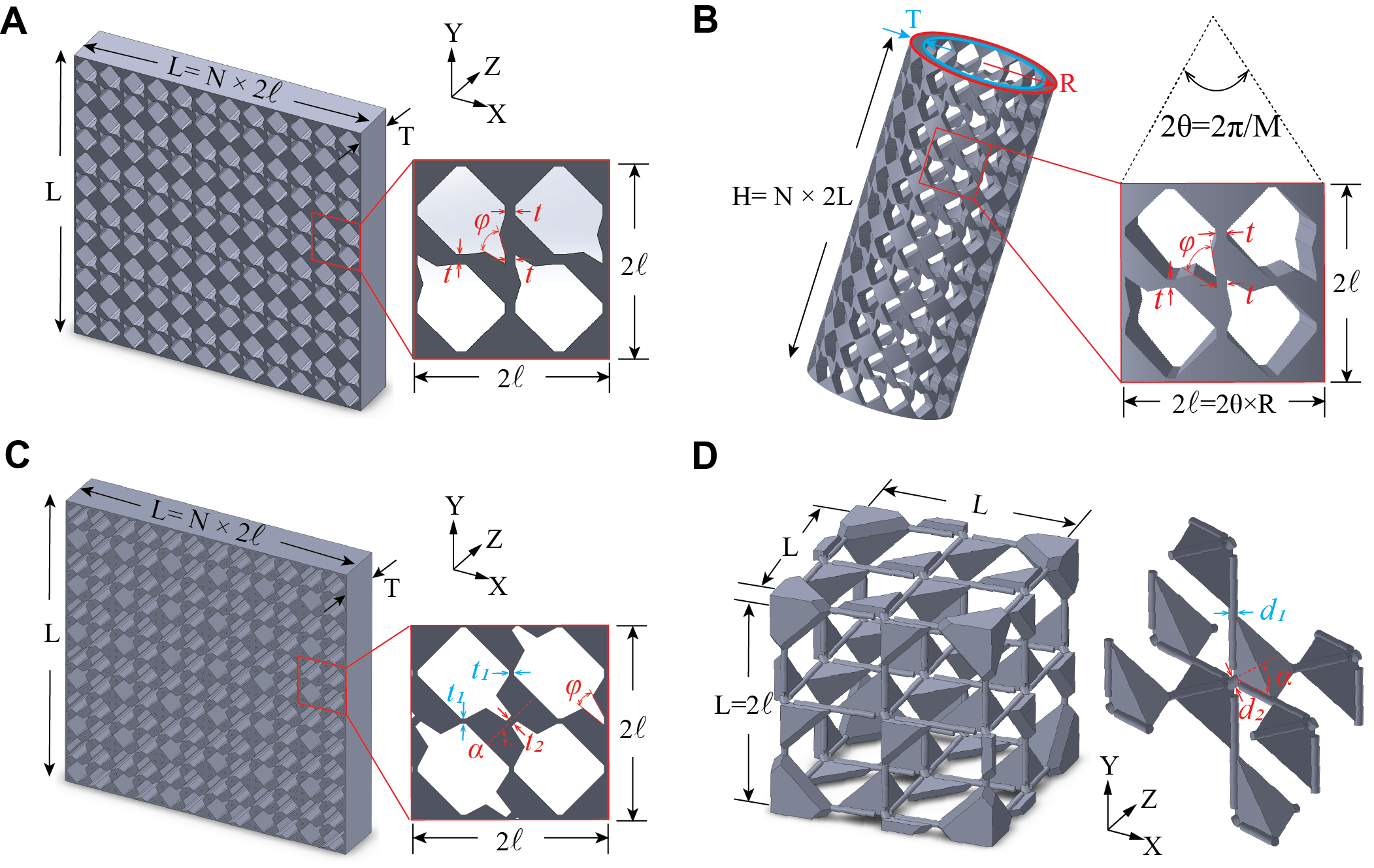}
\caption{\textbf{Geometry design of metamaterials with line modes.} \textbf{A,} a 2D metamaterial with one directional line-mode pattern. The unit cell has the size, $L$, ligament width, $t$, and contact angle $\phi$. \textbf{B,} A cylinder sample designed from the 2D line-mode pattern with a radius thickness, $T$, radius, $R$, and length, $H$. \textbf{C,} a 2D metamaterial with a 2D orthogonal line-mode pattern.The ligament in the cross-of-the-line modes has an angle $\alpha=45^{\circ}$ and width $t_{2}$, and the rest of the ligaments with the width $t_{1}$. \textbf{D,} A 3D structure with 3 orthogonal layer modes in X, Y, and Z directions, The ligament in the cross of the layer modes has an angle of $45^{\circ}$ and the diameter relation $d_{2}$,  and the rest of the ligaments have the width $t_{1}$. }
\label{fig:11}
\end{figure*}

The metamaterials we designed in this work comprise a plurality of unit cells arranged in a periodic pattern, wherein the unit cells comprise rigid parts connected by ligaments that can act as hinges during buckling. We create the following designs:

\begin{enumerate}
    \item Metamaterials with parallel line modes (Fig.~\ref{fig:11}A): we assemble the line-mode unit cell into a $N \times N$ metamaterial and extrude it to a thickness $T$. The metamaterial thickness $T$ is equal to the size of unit cell $\ell$ to avoid out-of-plane buckling under compression. All the ligaments in the loading path $Y$ have the same geometry size $t$ and orientation angle. The triangular raised parts between two connected ligaments are designed for the self-contact and the angle $\varphi$ is used to control the rotating angle and the compressible stroke. A smaller rotating angle can lead to less force drop in the post-buckling process, but sacrifice more compressible stroke. 
    \item Metacylinder (Fig.~\ref{fig:11}B): the 2D pattern with six line modes is projected to a cylinder to design a shock absorber with a continuous line-mode boundary condition. The metacylinder has an outer radius, $R$, height, $H$, and thickness, $T$, Compared to the 2D design, a much smaller ratio, $T/t \approx 5$, between the ligament width and thickness is needed to avoid the out-of-plane buckling.
    \item Metamaterials with orthogonal line modes (Fig.~\ref{fig:11}C): we changed two diagonal sub-units in the unit cell from the five-bar linkage to the six-bar linkage to have one more line mode in the orthogonal direction. To make the same buckling behavior in the two directions, we change the angle of the ligament in the middle of the unit cell from the vertical direction to $45^{\circ}$. The ligament with an angle has a smaller yielding condition under compression. A size ratio, $t_{1}/t_{2}=0.6$, between the vertical ligament and the $45^{\circ}$ ligament is used to ensure the same yielding condition. 
    \item 3D metamaterials with orthogonal layer modes (Fig.~\ref{fig:11}D): to expand the idea of line mode to a higher dimension, we designed a 3D unit cell consisting of tetrahedrons. Each tetrahedron is connected to the four tetrahedron neighbors by ligaments with a circular cross-section. A ligament diameter ratio $d_{1}/d_{2}=0.75$, between the vertical ligament and the $45^{\circ}$ ligament is used to ensure the same yielding condition.
    \item Crash Cans: A solid cylindrical shell is designed as a crash can with the same height $H$ as the metacylinder. The shell thickness is designed as 0.4 mm where the stiffness and strength of the crash can is close to the metacylinder with the ligament thickness $t=0.5$ mm. 
\end{enumerate}

\subsection{Sample fabrication}
\begin{figure*}[b!]
\centering
\includegraphics[width=0.4\linewidth]{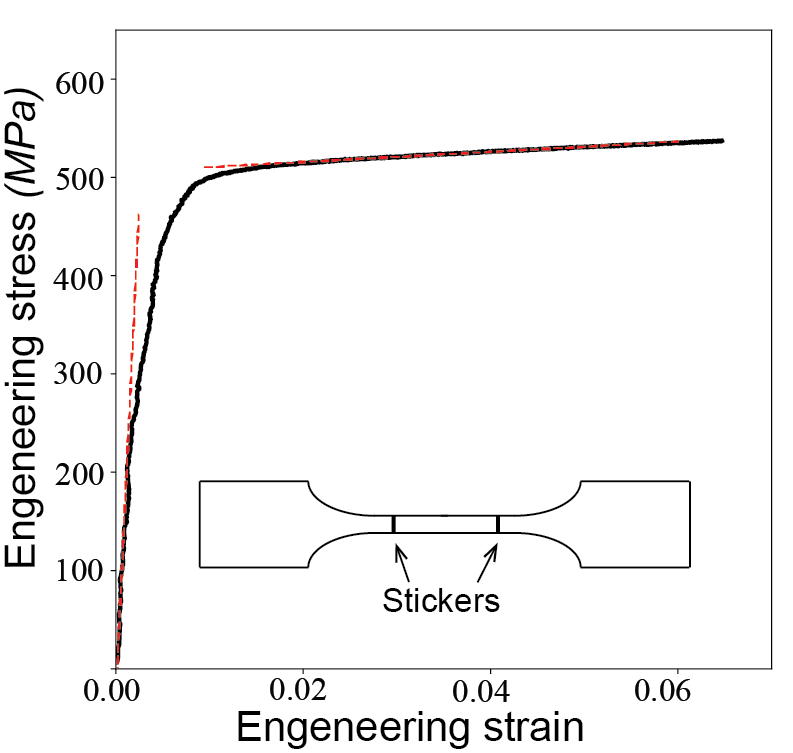}
\caption{Stress-Strain curve of 3D printed 316L steel where we linear fit the elastic modulus and tangent modulus. The fits provide measurements for the Young's modulus $E=200$ GPa, yield stress  $\sigma_y=500$ MPa and tangent modulus $E_t=500$ MPa.}
\label{fig:12}
\end{figure*}

\begin{figure*}[t!]
\centering
\includegraphics[width=0.9\linewidth]{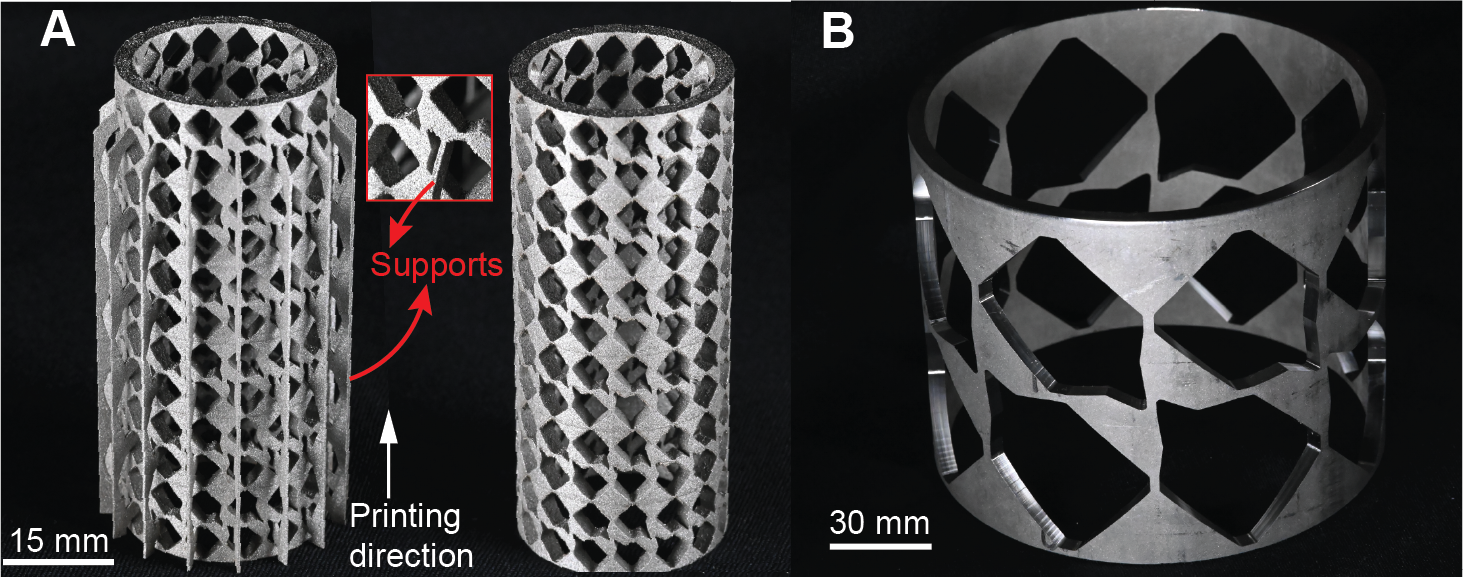}
\caption{\textbf{sample fabrication by additive manufacturing and milling} \textbf{A,} Additive manufactured metacylinder with six line modes. Supports are added at the overhang edges along the printing direction (left) and are removed after printing (right). \textbf{B,} Milled metacylinder with a single layer of line mode.}
\label{fig:13}
\end{figure*}

We used selective laser melting (SLM) technology (GE additive GlabR) to fabricate our metamaterials. The printing material we used is 316L stainless steel which has a very small ratio between tangent modulus ($E_{t}\approx500$  MPa) and elastic modulus ($E\approx200$ GPa), $E_{t}/E = 0.25 \%$,  and a relatively high yield stress $\sigma_{y} = 500$ MPa (see fig.~\ref{fig:12} for the calibration of elastoplastic properties). For the metamaterials with a 2D pattern, no supports are needed during the printing and we can cut the samples directly from the printing platform with a wire-cutting machine. We printed the metacylinder along the length direction with supports for the overhang edges of the blocks (fig.~\ref{fig:13}A). Such metacylinder can be also produced at the industrial scale without using additive manufacturing. We prove this by milling such a metamaterial pattern on a cylindrical tube with a diameter of 140 mm and a shell thickness of 4 mm (fig.~\ref{fig:13}B). 

\subsection{Mechanical tests}
The samples were tested under static and dynamic uniaxial compression. The static tests were processed in a universal testing machine (Instron 5985) equipped with a 300 kN load cell and compression plates, which enabled us to impose a compressive displacement with an accuracy of 0.01 mm and to record the force with $0.5\%$ of the reading accuracy. We first processed a continuous compressing test for the metacylinders with length ratio, $t/\ell=0.067,\; 0.084,\; 0.01$ in the displacement control with a speed of 0.2 mm/s (see Fig.~\ref{fig:4}AB). We then implemented a loading and unloading test for the metacylinder with $t/\ell=0.084$ where we used displacement control with a speed of 0.2 mm/s during the loading process and force control with the same speed during the unloading process until force-free. We repeat this process six times until all layers self-contact. 

For the dynamic test, we first tested at a 100 mm/s speed with a high-load hydraulic fatigue machine (MTS) equipped with a 300 kN load cell. For the higher loading speed, we used two types of drop towers to apply the impact to the metacylinder. The impact speed depends on the height of the free drop. The lab drop tower has a range of dropping height up to 1.75 m and of dropping weight from 5.5 kg to 22 kg. The maximum impact speed is up to 5.8 m/s. We conducted the multiple impact tests with an impact speed of 3.2 m/s and drop weight of 5.5 kg. For the single impact test, we applied an impact speed of 4.7 m/s and a drop weight of 15.5 kg The industry drop tower from Tata Steel has a range of dropping heights up to 10 m, and an impact speed is up to 14 m/s, which we used for the impact test at 10 m/s with a weight of 22.5 kg in this work.

\subsection{Imaging and data analysis}
The static tests were recorded using a high-resolution camera (Nikon D780 with 105 mm lens, resolution 1080 px $\times$ 1280 px, and frame rate 30 fps). The impact videos were recorded with a high-speed camera (Phantom 7510 and lens 105 mm with a frame rate of 126506 fps at 1024 px $\times$ 480 px), paired with two high-power LED-based front-light systems. White circle stickers were glued on the front of the dropping plate to facilitate position tracking. We then used Image-J software to track the position of the white stickers of each frame in the video. From the sticker's position at each frame, we calculate the speed, acceleration, and reaction force based on the drop weight and the noise-filtered displacement of the sticker.

\section{Additional results}

\subsection{Multiple impacts}
\begin{figure*}[t!]
\centering
\includegraphics[width=1.0\linewidth]{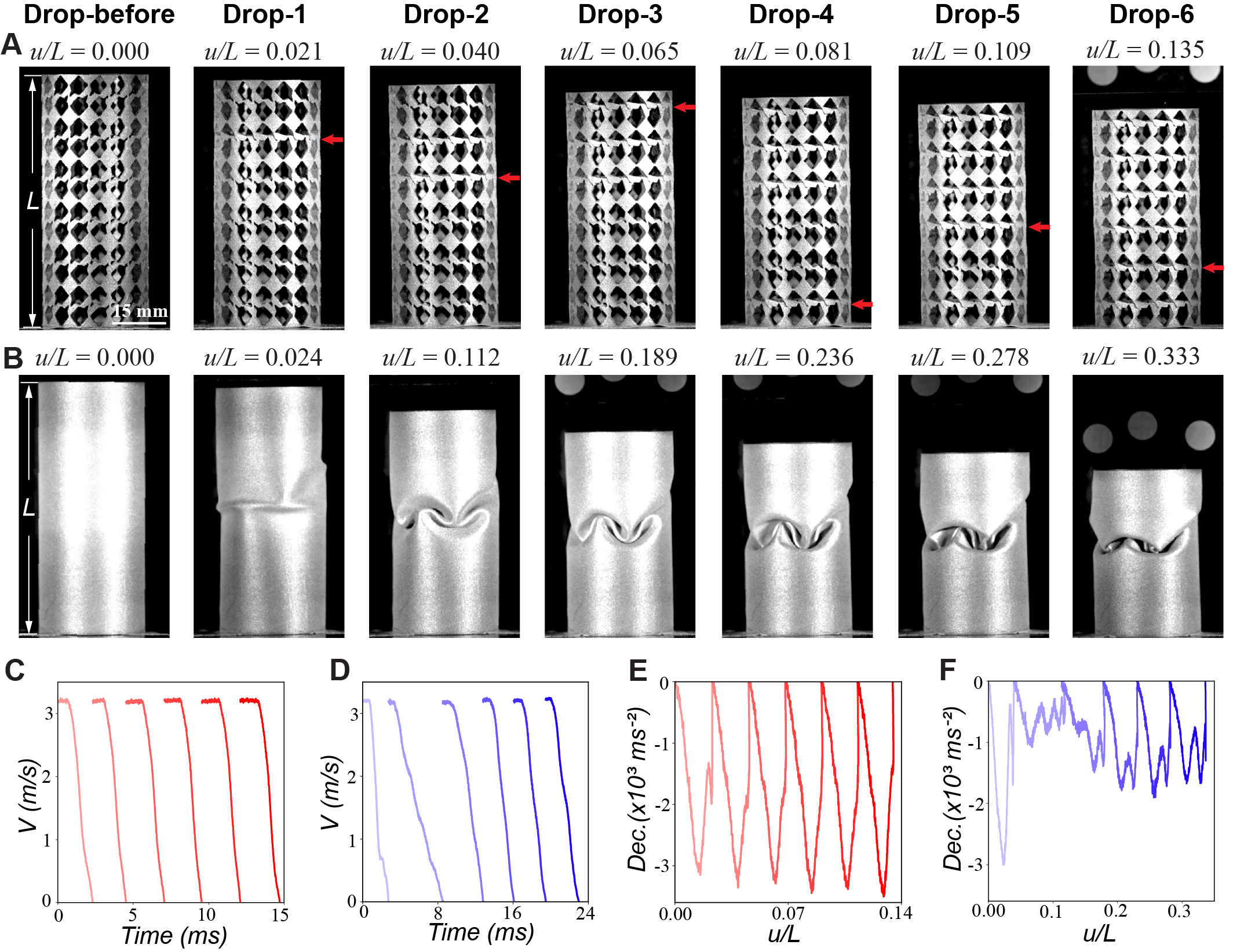}
\caption{\textbf{Multiple impacts.} \textbf{A, B, } snapshot comparisons at each impact of the metacylinder and the crash can with the same stiffness and strength. The speed of each 
 impact is $3.2$ m/s with a weight of $5.5$ kg. \textbf{ C, D,} the impact speed, $V$, vs. impact time curves at each impact of the metacylinder (C) and the crash can (D). \textbf{E, F,} The deceleration vs. compressive stroke under six distinct
dynamic impacts of the metacylinder (E) and crash can structures (F).}
\label{fig:14}
\end{figure*}

As we show in the Main Text, we can multiple times achieve the same initial properties after a static loading-unloading and dynamic multiple impact process by using the sequential layers of buckling and self-contact in the metamaterials (Fig.~\ref{fig:4}). Here we further strengthen this idea by comparing it to a crash can with the same stiffness and strength in a multiple-time impact test with the same impact speed $3.2\; m/s$ and drop weight $5.5 \; kg$. In the first impact of the metacylinder, one layer line mode buckles and gets into self-contact at a stroke of 0.021. The rest five line modes buckle in sequence in the next five impacts consuming almost the same stroke in each impact 0.135 stroke in total (Fig.~\ref{fig:14}A). In the first impact of crash can, the shell locally folds in the middle at a stroke of 0.024. In the second impact, the stroke highly increases to 0.112 with a further local folding at the same location. The rest of the impacts make the deformation localize further until a stroke of 0.333 (Fig.~\ref{fig:14}B). 
The uniformly increased stroke in the metacylinder results in the same deceleration in each impact (Fig.~\ref{fig:14}CE). In contrast, the deceleration of the first impact in the crash can is much higher compared to the rest of the impacts (Fig.~\ref{fig:14}DF). The stroke and deceleration difference between the first impact and the subsequent impacts in the crash can comes from the mechanism changing from stretch-dominated to bending-dominated after the first impact, which highly softens the structure. In contrast, the self-contact stiffening mechanism after each layer of buckling in the metacylinder can maintain the same stiffness and strength after each impact and absorb the same energy with the same stroke and deceleration.
\subsection{Out-of-axis compression}
\begin{figure*}[htb!]
\vspace{-3mm}
\centering
\includegraphics[width=1.0\linewidth]{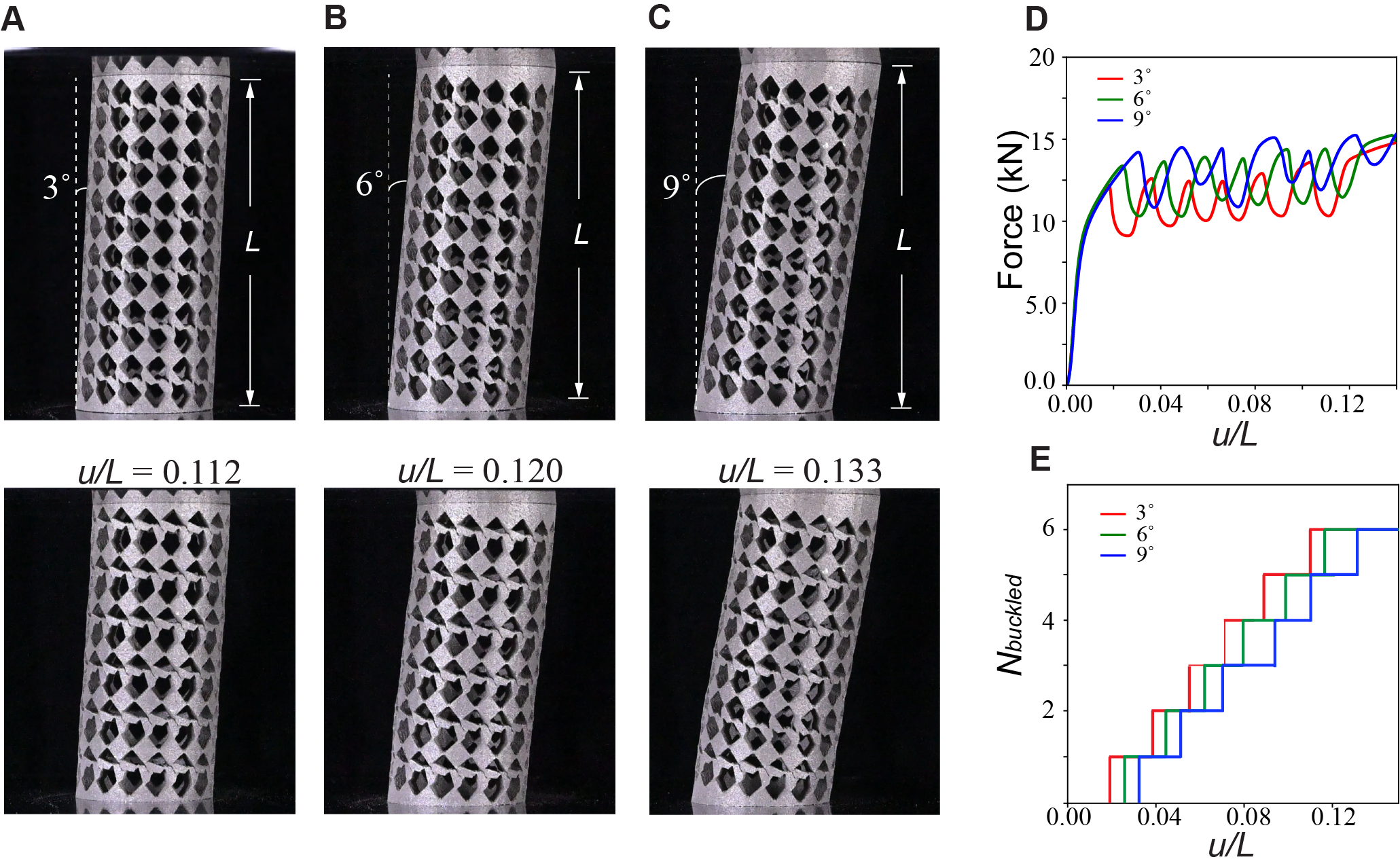}
\caption{\textbf{Out-of-axis compression of the metacylinders with six line modes} \textbf{A, B, C,} Compression of the metacylinders at three different oblique angles. \textbf{D,} the force $F$ and compression stroke $u/L$ of the the metacylinders with different angles. \textbf{D,} Number of buckled line modes $N_\textrm{buckled}$ vs. compressive stroke, $u/L$, for various out-of-axial angles.}
\label{fig:15}
\end{figure*}

We have already shown that the metacylinders have robust sequential yield buckling in different length ratios and a large range of loading speeds (Fig.~\ref{fig:4}B). Here, we further prove that this sequential mechanism is robust in metacylinders with different oblique angles, $3^{\circ}, 6^{\circ}$, and $9^{\circ}$ (Fig.~\ref{fig:15}ABC top). The length ratio $t/\ell$ and the number of line modes $N$ are the same as the metacylinder in Fig.~\ref{fig:4}A. Under compression, these three out-of-axis metacylinders all show similar sequential buckling behavior as the metacylinder tested orthogonal to the compression axis (Fig.~\ref{fig:15}ABC bottom and Fig.~\ref{fig:15}D).The buckling load (Fig.~\ref{fig:15}D) and buckling strain (Fig.~\ref{fig:15}E) of each line mode increase with oblique angle. With this experiment, we further confirm the robustness of the sequential yield buckling in metamaterials.

\subsection{Optimised metamaterials}
\begin{figure*}[htb!]
\centering
\includegraphics[width=1.0\linewidth]{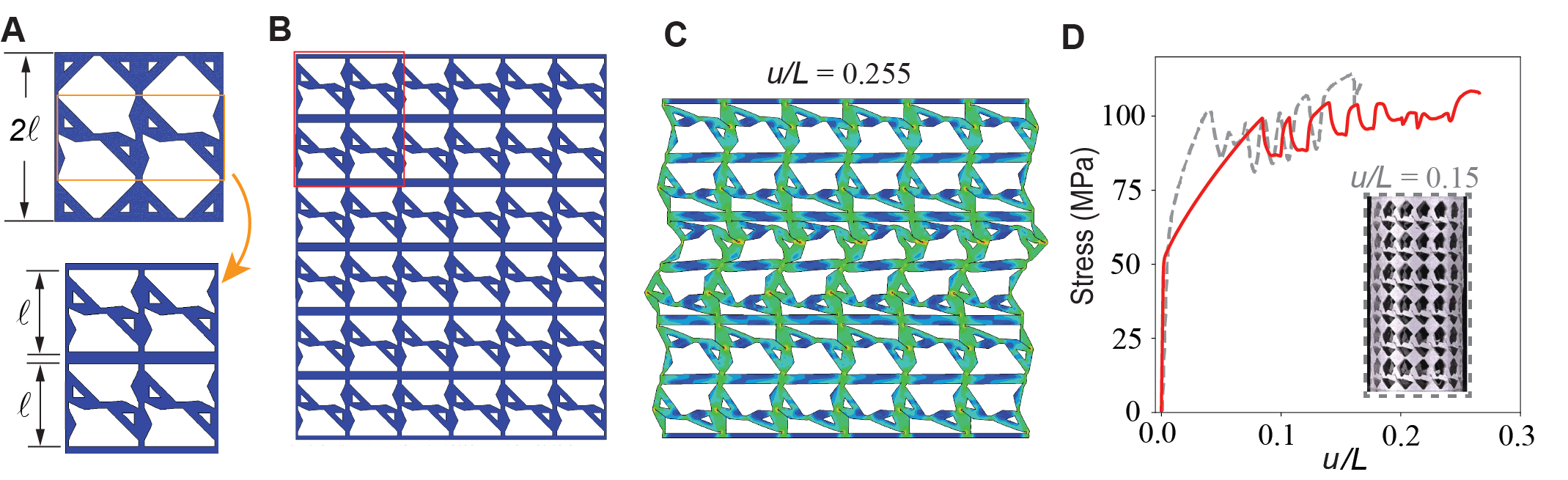}
\caption{\textbf{Geometry optimization.} \textbf{A,} A unit cell with single line mode (top) and the optimized unit cell with double line modes (bottom). \textbf{B,} A metamaterial consists of 6 $\times$ 6 unit cells.\textbf{C,} Snapshot of the optimized metamaterial with six line modes under compression. \textbf{D,} stress-strain curves of the metamaterial with double line-mode unit cells (red) and the metamaterial with single line-mode unit cells (grey, experimental result) .}
\label{fig:16}
\end{figure*}

We have proven that the metacylinders combine high stiffness and high energy absorption before $20\%$ of compressing stroke (Fig.~\ref{fig:4} and Fig.~\ref{fig:5}AB).
A higher number of buckling sequences and a larger compressible stroke with progressive energy absorption can be achieved through geometry optimization. To do this, we replace the top and bottom triangle blocks in the unit cell (Fig.~\ref{fig:16}A left) with two separating plates, which allows to design an optimized unit cell with two line modes (Fig.~\ref{fig:16}A). We then assemble such an optimized unit cell into a $3 \times 3$ metamaterial. Such metamaterial exhibits the same six-step sequential buckling, yet is stiffer and shows and double compressive line modes compared to the metamaterial with the same geometry size(Fig.~\ref{fig:16}CD).

\subsection{Multi-direction yield buckling}
\begin{figure*}[h!]
\centering
\includegraphics[width=1.0\linewidth]{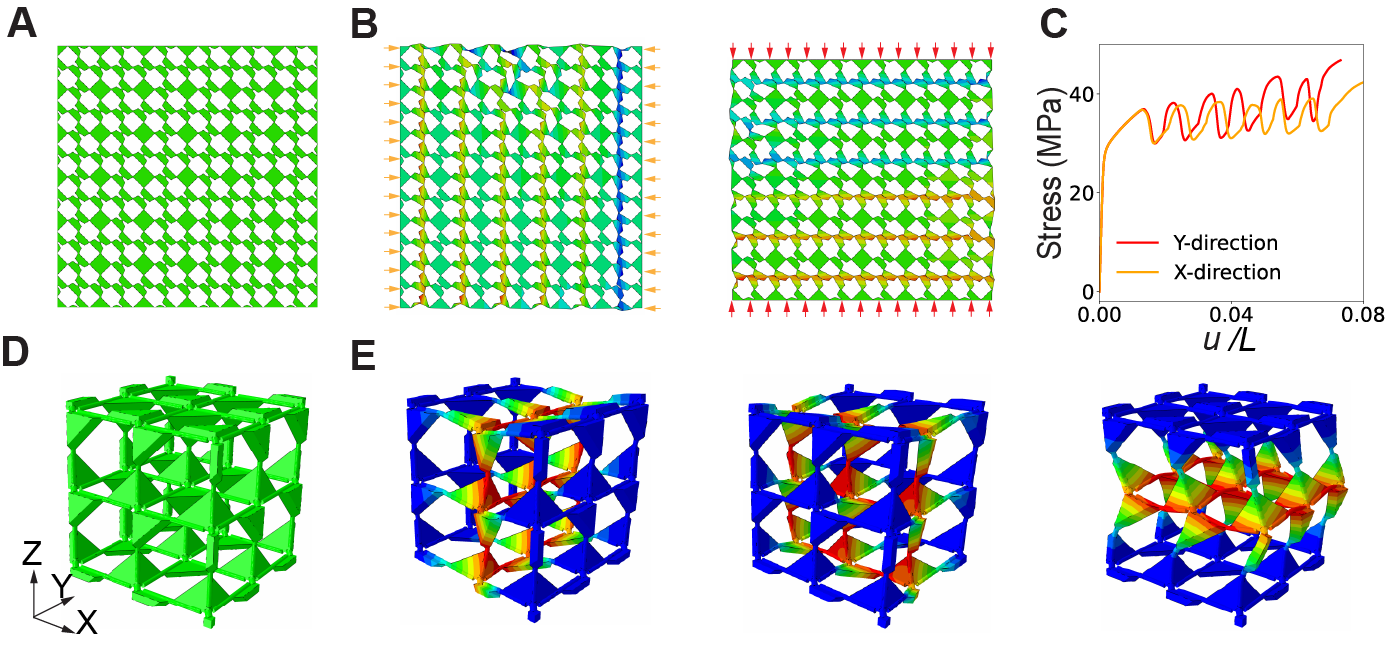}
\caption{\textbf{Yield buckling in multiple directions.} \textbf{ABC.} 
A metamaterial made from $6\times 6$ unit cells that exhibits line modes along the vertical and horizontal directions. (\textbf{AB}) Structure at rest (\textbf{A}) and under horizontal and vertical compression (\textbf{B}) in finite elements simulations (see also Supplementary Movie 4). 
(\textbf{C}) Corresponding force-displacement curve.
%\textbf{B,} the simulation shows the same 6 steps of localized buckling in X and Y directions. 
\textbf{DE.} 3D unit cell with 3 orthogonal buckling modes localized along surfaces. Structure (\textbf{D}) at rest and (\textbf{E}) in linear buckling stability analysis along three directions.}
\label{fig:17}
\end{figure*}

So far, we have proved that the metamaterials enabled by yield buckling perform as ideal shock absorbers, yet only along one direction. We expand the shock absorbers by designing orthogonal layers in two or three dimensions (Fig.~\ref{fig:17}A-E), which allow for enhanced shock absorption in two or three directions.
%buckling into two directions by using a 2D unit with two line modes. (Fig.~\ref{fig:5}A). 
In two dimensions, we modify the design of Fig.~\ref{fig:3}A to host line modes along two directions (Fig.~\ref{fig:17}). Our simulations confirm that the layers buckle in sequence when compressed in both directions respectively (Fig.~\ref{fig:17}B and Supplementary Movie 4). As a result, the force-displacement curves combine high stiffness prior to buckling and a wiggly plateau with 6 oscillations, inducing high dissipation (Fig.~\ref{fig:17}C).
%We then approve it with the simulation in a 2D metamaterial with $6 \times 6$ orthogonal line modes where the same 6 localized line modes behavior shows in X and Y directions (Fig.~\ref{fig:4}B). 
We generalize the concept to a 3D unit cell with three orthogonal layer modes (Fig.~\ref{fig:17}D and ~\ref{fig:11}D) and perform a mode analysis that confirms the existence of orthogonal layers in all three directions (Fig.~\ref{fig:17}E).

\subsection{Relative strength and stiffness of the metamaterials}
\begin{figure*}[h!]
\centering
\includegraphics[width=1.0\linewidth]{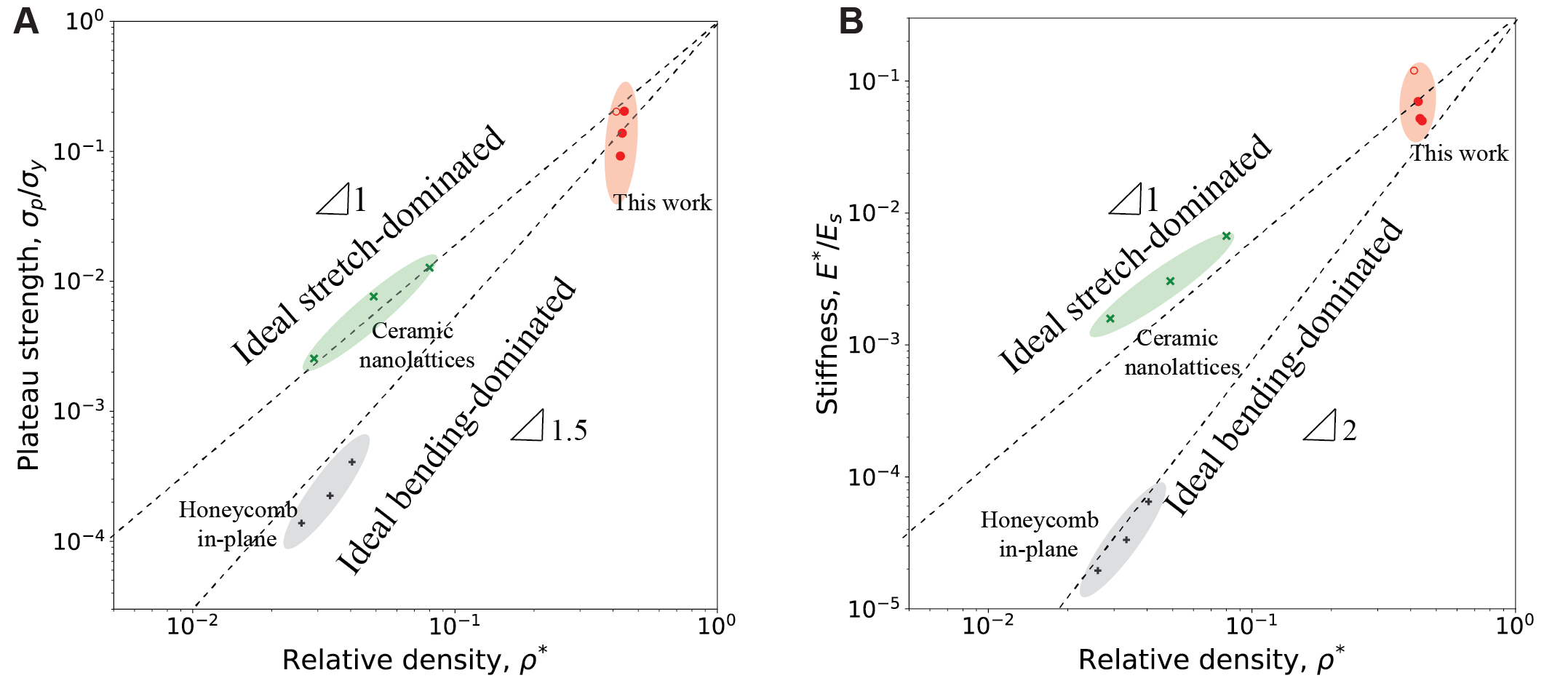}
\caption{\textbf{The relative plateau strength $\sigma_{pl}/\sigma_{y}$ and the relative stiffness $E/E_{s}$ vs. relative density of the metacylinders in this work compared with other two specific stretch-dominated and bending-dominated materials.}}
\label{fig:18}
\end{figure*}

We have shown the specific stiffness and specific energy absorption of the metacylinders compared to state-of-the-art competitive metamaterials (Fig.~\ref{fig:5}AB). Here, we further demonstrate the relative strength $\sigma_{pl}/\sigma_{y}$ (Fig.~\ref{fig:18}A) and relative stiffness $E^{*}/E_{s}$ (Fig.~\ref{fig:18}B) of the line-mode metamaterials changing with their relative density $\rho^{*}$ and the comparison with the other two classical stretching-dominated nanolattices and the bending-dominated honeycomb. Our line-mode metamaterials show much higher strength and stiffness increase in a very small change of relative density since the mechanical properties in the line-mode metamaterials are controlled by the size of ligaments rather than the global geometry change in the other two materials. 

\end{appendix}

\end{document}